\documentclass[aps,prb,preprint,superscriptaddress]{revtex4}

\usepackage{graphicx}
\usepackage{dcolumn}
\usepackage{pst-all}      

\newcounter{Alist}
\def\vxc{V^{\rm xc}}
\def\ei{\varepsilon_i}
\def\ej{\varepsilon_j}
\def\ef{E_F}

\newcommand{\bfq}{{\bf q}}
\newcommand{\bfk}{{\bf k}}
\newcommand{\bfr}{{\bf r}}

\newcommand{\bfR}{{\bf R}}

\newcommand{\val}{{\rm{VAL}}}
\newcommand{\core}{{\rm{CORE}}}
\newcommand{\itval}{{\it{val}}}
\newcommand{\itcore}{{\it{core}}}
\newcommand{\xc}{_{\rm{xc}}}

\newcommand{\req}[1]{Eq.~(\ref{#1})}

\def\ekn{{\varepsilon_{{\bf k}n}}}

\def\Ekn{{E_{{\bf k}n}}}
\def\Psikn{{\Psi_{{\bf k}n}}}

\def\rs{r^{\rm s}}

\begin{document}


\title{
Adequacy of Approximations in $GW$ Theory
}



\author{Mark van Schilfgaarde}
\affiliation{Arizona State University, Tempe, AZ, 85284}

\author{Takao Kotani}
\affiliation{Arizona State University, Tempe, AZ, 85284}

\author{Sergey V. Faleev}
\affiliation{Sandia National Laboratories, Livermore, CA 94551}

\date{\today}

\begin {abstract}

Following the usual procedure of the $GW$ approximation ($GW$A)
within the first-principles framework, we calculate the self energy
from eigenfunctions and eigenvalues generated by the local-density approximation (LDA).
We analyze several possible sources of error in the theory
and its implementation, using a recently development all-electron
method approach based on the full-potential linear muffin-tin orbital (LMTO) method.
First we present some analysis of convergence in some quasiparticle energies
with respect to the number of bands, and also their dependence on different
basis sets within the LMTO method.
We next present a new analysis of core contributions.
Then we apply the $GW$A to a variety of materials systems,
to test its range of validity.
For simple $sp$ semiconductors, $GW$A always underestimates bandgaps.
Better agreement with experiment is obtained when the
renormalization ($Z$) factor is not included, and we propose a justification for it.
We close with some analysis of difficulties in the usual $GW$A procedure.

\end{abstract}


\maketitle

\section{Introduction}

Even though the {\em GW} approximation ($GW$A) of Hedin\cite{hedin65} is as
old as the local-density approximation (LDA), it is still in its early
stages because of serious difficulties in its implementation.  In the usual
\emph{ab initio} procedure, $G$ and $W$ are constructed from the LDA
potential, which generate the self-energy $\Sigma=i G \times W$.
Additionally the quasiparticle energies (QPE) are usually approximated as a
perturbation correction to the LDA, from the matrix elements of the
diagonal parts of $\Sigma-V\xc^{\rm{LDA}}$; see Eqns.~\ref{eq:qpgwnc} and
~\ref{eq:qpgw} below.  In principle, it is well-defined as a procedure.
However, there is a controversy what the numerical result of this procedure
is for semiconductors.  Nearly always the $GW$A is implemented in
conjunction with the pseudopotential (PP) approximation, which we will call
PP$GW$.  It was widely thought that PP$GW$ predicts bandgaps in
semiconductors to rather high accuracy.  However, recent all-electron {\em
GW} calculations to survey bandgaps in semiconductors using the
full-potential LMTO (FP-LMTO) by Kotani and van Schilfgaarde
\cite{kotani02} result in bandgaps that are generally smaller than
experimental values.  The result is confirmed by other calculations using
two independently-developed full-potential linear augmented plane wave
(FP-LAPW) codes: one by Usuda, Hamada, Kotani and van
Schilfgaarde\cite{Usuda02} and another by Friedrich, Schindlmayr,
Bl{\"u}gel and Kotani \cite{Christoph06}.  These methods all use
essentially the same $GW$ codes originally developed in conjunction with
the LMTO method \cite{kotani02}; they differ only in the input
eigenfunctions.  Calculations from other, independently developed
all-electron $GW$ methods \cite{hamada90,weiku02,alouani03} are consistent
with this conclusion\cite{ldacorenote}.

Tiago, Ismail-Beigi, and Louie\cite{tiago04} used the PP$GW$ scheme that
included Si $2s$ and $2p$ cores in the valence to analyze the dependence of
some semiconductor bandgaps on the number of unoccupied states $N'$ used to
construct the self-energy.  They suggested that the discrepancy between
all-electron $GW$ and PP$GW$ gaps could be attributed to incomplete convergence in
the all-electron calculations.  To address this point, the convergence in
$N'$ is taken up in Sec.~\ref{sec:nconvergence}.  We begin with an outline
our all-electron $GW$ method (Sec.~\ref{sec:methodology}); it includes a
comparison of the energy bands in Si to those of an APW calculation
taken from Friedrich et al.~\cite{Christoph06},
establish the method's ability to reproduce near-exact LDA eigenvalues.  In
Sec.~\ref{sec:basisconvergence} we show how selected QPEs change with
increasingly larger LMTO basis sets for a variety of semiconductors.  The
results are weakly dependent on basis even for relatively small basis
sets.  We present some rationale for why this should be so, and note the
implications for both precision and efficiency in implementations of the
$GW$A for basis sets in general.

Because our results are well converged for either kind of test, we still
think that PP$GW$ is problematic, in contradistinction to the conclusions
in a recent paper by Delaney, Garc\'ia-Gonzal\'ez, Rubio, Rinke and Godby
\cite{delaney04}, who showed that all-electron $GW$ and PP$GW$ give
essentially the same result for the Be free atom.  However, the Be atom is
a special case in part because the all-electron and pseudo radial functions
should closely correspond to each other (the $2p$ radial function has no
nodes, and the only core that is orthogonalized or pseudized is the deep $1s$
core ($\epsilon^{\rm{LDA}}\approx-105~$eV); moreover the PP is constructed
with the atom itself as reference.
Pseudopotentials are constructed to solve LDA reliably, but not to solve
the $GW$A.  There are now many detailed checks comparing PP-LDA results
against the corresponding all-electron values; but there are few similar
comparisons for $GW$.  The discrepancies between all-electron and PP$GW$
appear to be much smaller when PP$GW$ includes the highest lying core
states in the valence.

Sec.~\ref{sec:core} analyzes different core contributions to the
QPE in several semiconductors.  This provides some insight as to
what approximations may be made concerning the core; we also
briefly consider some aspects of PP$GW$ in this context.

In Sec.~\ref{sec:gldawlda} we show some new results for a variety of
materials, as well as repeating some previously reported
calculations\cite{kotani02,faleev04,qp06,chantis06a,gapnote} with rather
tight tolerances.  We confirm that the usual $GW$A procedure generally
underestimates bandgaps.  We also show that a partial self-consistency can
be accomplished by calculating QPEs without the renormalization factor
$Z$ (i.e.  $Z$=1).  Semiconductor bandgaps are systematically improved
using $Z$=1, though they continue to be underestimated.  An important
reason for this is that the LDA overestimates the screening of $W$,
resulting in an underestimate of $\Sigma=iGW$ and bandgaps.  We show that
the adequacy of the $GW$A varies from system to system: only when the
starting LDA is reasonably good does the $GW$A reasonably predict QP
energies.  Thus, some kind of self-consistency is necessary to obtain
reasonable results for wide range of materials.\cite{qp06}

\section{Methodology}
\label{sec:methodology}

\subsection*{All-electron LDA in FP-LMTO}
\label{subsec:LDA}

Before turning to the analysis, we briefly describe the LDA method we use
as input for the $GW$ calculations.  (Readers interested in conclusions of
this paper \emph{not} related to basis-set issues can skip this section).
An early version of this method was presented in Ref.~\onlinecite{lmfchap};
we describe here how additional local orbitals are included to extend the
linear method.  Local orbitals are essential to the analysis, because QPE
in $GW$A are sensitive to a wider range of states than in LDA (e.g., the
LDA depends only on occupied states).  One consequence is that the linear
approximation inherent in standard linear and pseudopotential methods is
less reliable for the $GW$A than for the LDA.  The basis functions used in
the present technique are a generalization\cite{lmfchap} of the
standard\cite{Andersen75} method of linear muffin-tin orbitals (LMTO).
Conventional LMTOs consist of atom-centered envelope functions augmented
around atomic sites by a linear combination of radial wave functions
$\varphi$ and their energy derivatives $\dot\varphi$.
$\varphi=\varphi_{Rl}({\varepsilon_{Rl},r})$ is the solution of radial
Schr\"odinger equation at site $R$ at some linearization energy
$\varepsilon_{Rl}$.  A linear method matches the $\{\varphi,\dot\varphi\}$
pair to value and slope of the envelope function at each augmentation
sphere boundary, which means that the LDA Schr\"odinger equation can be
solved more or less exactly to first order in
$\varepsilon-\varepsilon_{Rl}$ inside each augmentation sphere.  Envelope
functions in the standard LMTO method consist of Hankel functions.  In the
present basis\cite{lmfchap} the envelope functions are smooth, nonsingular
generalizations\cite{Bott98} of the Hankel functions: the $l$=0 smooth
Hankel satisfies the equation
\begin{eqnarray}
 (\nabla^2+\varepsilon) {H}_0(\varepsilon,\rs;\bfr) &=& -4 \pi g_0(\rs,r)\\
g_0(\rs;r) &=& \left(\sqrt{\pi}\rs\right)^{-3} \exp\left(-(r/\rs)^2\right)
           \quad \to \delta(\bfr) \hbox{ as}\  \rs\to 0 \nonumber
\label{eq:defsmh0}
\end{eqnarray}
and reduces to a usual Hankel function in the limit $\rs\to 0$.
$H_L$ for higher $L$=($l$,$m$) are obtained by
recursion\cite{Bott98}.  The basis can be divided into three types of functions:
\setcounter{Alist}{0}
\begin{list}{({\it \roman{Alist}})\,}{\leftmargin 0pt \itemindent 24pt \usecounter{Alist}\addtocounter{Alist}{0}}

\item A muffin-tin orbital (MTO) $\chi_{RjL}$ which consists of a smoothed Hankel
  centered at nucleus $\bfR$ and augmented by linear combinations of
  $\varphi_{Rl}$ and $\dot\varphi_{Rl}$ for each $L$ channel inside every
  augmentation sphere
\begin{equation}
\chi_{RjL}(\bfr) = {H}_L(\varepsilon_{Rjl},\rs_{Rjl};\bfr-\bfR)
                 + \sum\limits_{R'k'L'} {C_{R'k'L'}^{RjL} \{ \tilde P_{R'k'L'} ({\bfr}) - P_{R'k'L'} ({\bfr})\} }
\label{eq:defmt}
\end{equation}
$P_{R'k'L'}$ is a one-center expansion of
${H}_L(\varepsilon_{Rjl},\rs_{Rjl};\bfr-\bfR)$, and $\tilde P_{R'k'L'}$ is
a linear combination of $\varphi_{Rl}({\varepsilon_{Rjl},r})$ and
$\dot\varphi_{Rl}({\varepsilon_{Rjl},r})$ that matches $P_{R'k'L'}$ at the
augmentation sphere radius.  Expansion coefficients $C_{R'k'L'}^{RjL}$ are
chosen to make $\chi_{RjL}(\bfr)$ smooth across each augmentation boundary.
$\varepsilon_{Rjl}$ is chosen to be at or near the center of the occupied
part of that particular $l$ channel.  Products of such functions enters
into the construction of the hamiltonian and output density.  The present method
differs in significant ways from the usual LMTO and LAPW methods, and is
not a simple product of MTO Eq.~(\ref{eq:defmt}), and bears
some resemblance to the Projector Augmented Wave prescription\cite{PAW}.
It greatly facilitates $l$ convergence in the augmentation; see
Ref.~\onlinecite{lmfchap}.

\item ``Floating orbitals'' consisting of the same kind of function as
  ($i$), but not centered at a nucleus.  Thus, there is no augmentation
  sphere where the envelope function is centered.  There is no fundamental
  distinction between this kind of function and the first type, except
  that the distinction is useful when analyzing convergence.  Floating
  orbitals make little difference in LDA calculations; but a basis
  consisting of purely atom centered envelope functions is not quite
  sufficient to precisely represent the interstitial over the wide energy
  window needed for $GW$ calculations.  Without their inclusion, errors in
  QPE of order 0.1~eV cannot be avoided, as we will show.

\item A kind of ``local orbital'' which has a structure similar to ($i$).
  The fundamental distinction is that the ``head'' (site where the envelope
  is centered) consists of a new radial function $\phi^z_l$ evaluated at
  energy $\varepsilon^z_l$ either far above or far below the linearization
  energy ${\varepsilon_l}$.  For deep, core-like orbitals $\phi^z_l$ is
  integrated at the core energy; for high-lying orbitals $\varepsilon^z_l$
  is typically taken 1-2 Ry above the Fermi level $\ef$.  In the
  former case a tail is attached, with its smoothing radius $\rs$ chosen to
  make the kinetic energy continuous across the ``head'' augmentation
  sphere. It is thus atypical of conventional local orbitals, as it is
  nearly an eigenstate of the LDA hamiltonian without requiring other
  basis functions.

\end{list}
As is well known, the reason for using augmented wave methods in general
(and especially the LMTO method), is that the Hilbert space of
eigenfunctions in the energy range of interest is spanned by much fewer
basis functions than with other basis sets.  In the present method two
envelope functions $j=1,2$ are typically used for low $l$ channels $s$,
$p$, and $d$, and one for higher $l$ channels ($f$ and sometimes $g$).  The
augmentation + local orbital procedure ensures that the basis is reasonably
complete inside each augmentation sphere within a certain energy window;
the envelope functions + floating orbitals ensures completeness of the
basis in the interstitial.  The distinction between standard LMTO envelope
functions and the smoothed ones used here is important because the
generalized form significantly improves this convergence.  Core states not
treated as local orbitals are handled by integrating the radial
Schr\"odinger equation inside an augmentation sphere and attaching a
smoothed Hankel tail, allowing it to spill into the interstitial.  Thus the
Hartree and exchange correlation potentials are properly included; only the
matrix element coupling core and valence states is neglected.

\begin{figure}[!htbp]
\includegraphics[angle=0,width=.45\textwidth,clip]{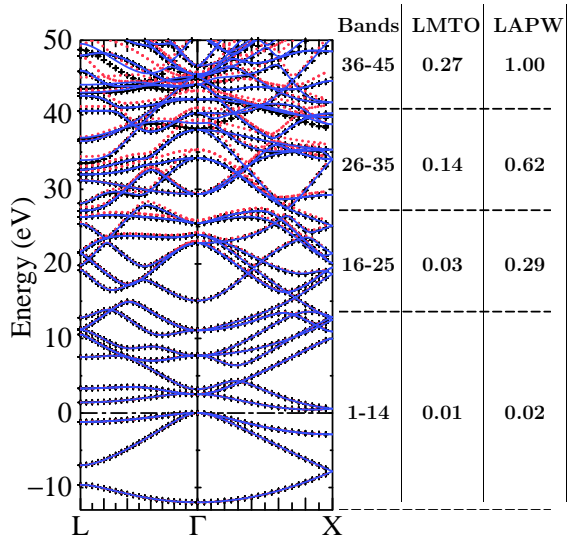}
\caption{LDA energy bands in Si, computed by different methods.  APW bands
  from Ref.~\protect\onlinecite{apwnote} are denoted by ``{\bf{}+}'' and
  can be regarded as near exact.  Dots denote bands calculated by the same
  authors using the LAPW method, without local orbitals.  Solid lines
  denote bands computed by the present generalized LMTO method, including
  local and floating orbitals as described in the text (180 energy bands
  were included in the basis).  On the right is a table of the RMS
  deviation relative to the APW bands for several energy windows, computed
  along the L$-\Gamma$ line.  First column denotes the range of bands used
  to compute the RMS deviation; the horizontal dashed lines denote the
  approximate energy window for each range.  Second and third columns
  denote the how the present method and the LAPW method respectively deviate
  from the APW bands, in eV, as described in the text.}
\label{fig:sildabands}
\end{figure}

When used in conjunction with $GW$ calculations we typically add local
orbitals for states not spanned by $\{\phi,\dot\phi\}$, and whose center of
gravity falls within $\sim\pm 25$~eV of the Fermi level $\ef$.  Both the
low-lying and high-lying states can be important, and we shall return to it
later.  Fig.~\ref{fig:sildabands} shows the effects of linearization in Si,
where an APW calculation of the LDA energy bands is
available\cite{apwnote}.  Friedrich et al.~\cite{apwnote} compared LDA
bands generated by a full APW calculation to those generated by LAPW.  They
are reproduced in Fig.~\ref{fig:sildabands}, together with the bands
generated by the LMTO+local+floating orbitals described above.  The LAPW
and APW are nearly indistinguishable on the scale of figure for energies up
to $\sim$25 eV.  For energies above 25 eV, the LAPW begins to deviate from
the other two, showing the effects of linearization.  The APW and
generalized LMTO bands are essentially indistinguishable on the scale of
figure for energies up to $\sim$40 eV.  Above that, slight differences
begin to appear; the differences gradually increase for still higher
energies.  Fig.~\ref{fig:sildabands} also tabulates the RMS deviation from
the APW bands for several energy windows.  The present method agrees with
the APW bands to $\sim$0.01 eV for levels within $\ef\pm{}1$Ry, and to
$\sim$0.25 eV for levels below $\ef+4$Ry.  Friedrich et al. report similar
improvements to their LAPW bands when local orbitals are
added\cite{Christoph06}.  They also compare bands generated by a PP, and
show that the errors are comparable to the conventional LAPW method.
Fig.~\ref{fig:sildabands} establishes rather convincingly that the present
method is nearly complete over a rather wide energy window in Si.  When
local orbitals are included, it is comparable to an LAPW method that
includes local orbitals\cite{apwnote}, and it is superior both to PP and
conventional LAPW methods.

\subsection*{All-electron $GW$ with mixed basis for $W$}
\label{subsec:allgw}

We briefly describe our all-electron implementation of the $GW$
approximation.  A more detailed account will be given
elsewhere\cite{longpap}.  The self-energy $\Sigma$ is
\begin{equation}
\Sigma(\bfr, \bfr', \omega)
=\frac{i}{2\pi}\int d\omega' G\left(\bfr,\bfr',\omega-\omega'\right)e^{i\delta\omega'} W\left(\bfr,\bfr',\omega'\right).
\label{eq:defsig}
\end{equation}
In this paper $G$ will be taken to be the one-body
non-interacting Green function as computed by the LDA,
and the screened Coulomb interaction $W$ is computed in the
random-phase approximation (RPA) from $G$.  Both $G$ and $W$ are obtained from the
LDA eigenvalues $\ekn$ and eigenfunctions $\Psikn$.  For a periodic
hamiltonian, we can restrict $\bfr$ and $\bfr'$ to a unit cell and write
$G$ as
\begin{equation}
G_{\bf{k}}\left( \bfr,\bfr',\omega \right)
= \sum\limits_{n}^{\rm{All}} {
\frac{\Psikn(\bfr) \Psikn^*(\bfr')}
     {{\omega -\ekn \pm i\delta }}}.
\label{eq:defg}
\end{equation}
The infinitesimal $-i\delta$ is to be used for occupied states, and
$+i\delta$ for unoccupied states.  $W$ is written as
\begin{equation}
W = \epsilon^{-1} v = \left(1-v \Pi\right)^{-1} v
\label{eq:defw}
\end{equation}
where $\Pi=-iG\times{}G$ is the bare polarization function shown below,
$v ={e^2}/{|\bfr-\bfr'|}$ is the bare coulomb interaction, and $\epsilon$
is the dielectric function.  For simplicity, the spin degree of freedom is
omitted.

Neglecting the off-diagonal part of $\Sigma$, we can evaluate QPE $\Ekn$ from
\begin{equation}
\Ekn=\ekn + Z_{\bfk n}[\langle\Psi_{\bfk n}|
\Sigma(\bfr,\bfr',\ekn)|\Psikn \rangle
- \langle\Psi_{\bfk n}|V_{\rm xc}^{\rm{LDA}}(\bfr)|\Psikn \rangle].
\label{eq:e1shot}
\end{equation}
$Z_{{\bfk}n}$ is the quasi-particle (QP) renormalization factor
\begin{equation}
Z_{{\bf k}n}=\left[ 1-\langle\Psikn|
\frac{\partial}{\partial\omega} \Sigma({\bf r},{\bf r}^{\prime},
\ekn)|\Psikn \rangle \right]^{-1},
\label{eq:defzfac}
\end{equation}
and accounts for the fact that $\Sigma$ is evaluated at the LDA energy
rather than at the QPE.  Eq.~(\ref{eq:e1shot}) is the customary way QPEs are
evaluated in $GW$ calculations.  In Sec.~\ref{sec:gldawlda}, we present an
argument that using $Z$=1 (or neglecting the $Z$ factor) is a better choice
than Eq.~(\ref{eq:defzfac}), and how the QPE is affected in actual
calculations.  However, the results presented here use the $Z$ factor
except where noted.

In FP-LMTO, eigenfunctions of the valence states are expanded in linear
combinations of Bloch summed MTOs, Eq.~(\ref{eq:defmt})
\begin{eqnarray}
\label{lmtopsi}
\Psikn(\bfr) = \sum_{RjL} a^n_{RjL} \chi^{\bfk}_{RjL}({\bf r}),
\end{eqnarray}
Inside augmentation sphere $\bfR$, the Hilbert space of the valence
eigenfunction $\Psikn(\bfr)$ consists of the pair (or triplet) of orbitals
($\varphi_{Rl}$, $\dot\varphi_{Rl}$ or $\varphi_{Rl}$, $\dot\varphi_{Rl}$,
$\varphi^z_{Rl}$) at that site\cite{ltrunc}, and can be represented in a
compact notation $\{\varphi_{Ru}\}$. $u$ is a compound index for both $L$
and one of the $(\varphi_{Rl}$, $\dot\varphi_{Rl}$, $\varphi^z_{Rl}$)
triplet.  The interstitial is comprised of linear combinations of envelope
functions consisting of smooth Hankel functions, which can be expanded in
terms of plane waves\cite{Bott98}.  Therefore the $\Psikn(\bfr)$ can be
written as a sum of augmentation and interstitial parts
\begin{eqnarray}
\Psikn(\bfr)
= \sum_{Ru}    \alpha^{{\bfk}n}_{Ru} \varphi^{\bf k}_{Ru}({\bf r})
 + \sum_{\bf G}  \beta^{{\bfk}n}_{\bf G} P^{\bf k}_{\bf G}({\bf r}),
\label{def:psiexp}
\end{eqnarray}
where the interstitial plane wave (IPW) is defined as
\begin{eqnarray}
P^{\bf k}_{\bf G}({\bf r}) &=& 0  \ \ \ {\rm \ if \ {\bf r} \in any \ MT} \nonumber \\
        &=&   \exp (i ({\bf k+G}) {\bf r}) \ \ \ {\rm otherwise},
\end{eqnarray}
and the $\varphi^{\bf k}_{R u}$ are Bloch sums of $\varphi_{R u}$
\begin{eqnarray}
\varphi^{\bf k}_{R u}({\bf r}) &\equiv& \sum_{\bf T} \varphi_{R u}({\bf r-R-T}) \exp(i {\bf k\cdot{}T}).
\end{eqnarray}
{\bf T} and {\bf G} are lattice translation vectors in real and reciprocal
space, respectively.

Throughout this paper, we will designate eigenfunctions constructed from MTOs
as ``\val''.  Below them are the core eigenfunctions which we designate as
``\core''.   There are two fundamental distinctions between \val\ and \core:
eigenfunctions: the latter are constructed independently by integration of
the spherical part of the LDA potential, and they not included in the
secular matrix.  Second, the cores are confined to MT spheres\cite{coretrunc}.
\core\ eigenfunctions are also expanded using \req{def:psiexp} in a trivial
manner; $\beta^{{\bfk}n}_{\bf G}=0$ and only one of
$\alpha^{{\bfk}n}_{Ru}\ne0$.  The discussion below applies to all
eigenfunctions, \val\ and \core.

Through Eq.~(\ref{def:psiexp}), products $\Psi_{{\bf k_1}n}\times\Psi_{{\bf
k_2}n'}$ can be expanded by $P^{\bf k_1+k_2}_{\bf G}({\bf r})$ in the
interstitial region because $P^{\bf k_1}_{\bf G_1}({\bf r}) \times P^{\bf
k_2}_{\bf G_2}({\bf r})= P^{\bf k_1+k_2}_{\bf G_1+G_2}({\bf r})$.  Within
sphere $R$, wave function products can be expanded by
$B_{Rm}^{\bf{k_1+k_2}}({\bf r})$, which is the
Bloch sum of the product basis $\{B_{Rm}({\bf r})\}$, which in turn is
constructed from the set of products
$\{\varphi_{Ru}({\bfr})\times\varphi_{Ru'}({\bfr})\}$ adapting the
procedure by Aryasetiawan\cite{prodbasis94}.
Eq.~(\ref{def:psiexp}) is equally valid in a LMTO or LAPW framework, and
eigenfunctions from both types of methods have been used in this $GW$
scheme\cite{Usuda02,Christoph06}.  We restrict ourselves to
LMTO-derived basis functions here.

We define the mixed basis
$\{M^{\bf k}_I({\bf r}) \}\equiv \{ P^{\bf k}_{\bf G}({\bf r}), B_{Rm}^{\bf k}({\bf r})\}$,
where the index $I\equiv \{ {\bf G},Rm\}$ classifies the members of the basis.
By construction, $M^{\bf k}_I$ is a good basis set for the expansion of
products of $\Psikn$.  Complete information to calculate $\Sigma$ and $E_n({\bf k})$ are
matrix elements of the products $\langle \Psi_{{\bf q}n}| \Psi_{{\bf q-k}n'} M^{\bf k}_I \rangle$,
the LDA eigenvalues $\varepsilon_{{\bf k}n}$,
the Coulomb matrix $v_{IJ}({\bf k})
\equiv \langle M^{\bf k}_{I} |v|  M^{\bf k}_{J} \rangle$,
and the overlap matrix $\langle M^{\bf k}_I |  M^{\bf k}_{J} \rangle$.
(The overlap matrix of IPW is necessary because
 $\langle P^{\bf k}_{\bf G} |  P^{\bf k}_{\bf G'} \rangle \ne 0$
for ${\bf G} \ne {\bf G}'$.)
The Coulomb interaction is expanded as
\begin{eqnarray}
\label{coulombexpand}
v({\bf r},{\bf r}') &=&
\sum_{{\bf k},I,J} |\tilde{M}^{\bf k}_I \rangle
v_{IJ}({\bf k}) \langle \tilde{M}^{\bf k}_J|,
\end{eqnarray}
where we define
\begin{eqnarray}
&& |\tilde{M}^{\bf k}_{I} \rangle \equiv \sum_{I'}
   |M^{\bf k}_{I'} \rangle (O^{\bf k})^{-1}_{I'I} \, , \\
&& O^{\bf k}_{I'I} = \langle M^{\bf k}_{I'} |  M^{\bf k}_I \rangle.
\end{eqnarray}
$W$ and the polarization function $\Pi$ shown below are expanded
in the same manner as Eq.~(\ref{coulombexpand}).

The exchange part of $\Sigma$ is written in the mixed basis as
\begin{eqnarray}
\langle \Psi_{{\bf q}n}|\Sigma_{\rm x} |\Psi_{{\bf q}n} \rangle
&&=\sum^{\rm BZ}_{{\bf k}}  \sum^{\rm  occ}_{n'}
\langle \Psi_{{\bf q}n}| \Psi_{{\bf q-k}n'} \tilde{M}^{\bf k}_I \rangle
v_{IJ}({\bf k})
\langle \tilde{M}^{\bf k}_J \Psi_{{\bf q-k}n'} | \Psi_{{\bf q}n} \rangle.
\label{eq:sigx}
\end{eqnarray}

The screened Coulomb interaction $W_{IJ}({\bf q},\omega)$ is calculated
through Eq.~(\ref{eq:defw}), where the polarization function $\Pi$ is written
\begin{eqnarray}
\Pi_{IJ}({\bf q},\omega)
&&=\sum^{\rm BZ}_{{\bf k}}  \sum^{\rm  occ}_{n} \sum^{\rm  unocc}_{n'}
\langle \tilde{M}^{\bf q}_I \Psi_{{\bf k}n} |\Psi_{{\bf q+k}n'} \rangle
\langle \Psi_{{\bf q+k}n'}| \Psi_{{\bf k}n} \tilde{M}^{\bf q}_J \rangle \nonumber \\
&& \times
\left(\frac{1}{\omega-\varepsilon_{{\bf q+k}n'}+\varepsilon_{{\bf k}n}+i \delta}
-\frac{1}{\omega+\varepsilon_{{\bf q+k}n'}-\varepsilon_{{\bf k}n}-i \delta}\right). \label{dieele}
\label{eq:polf}
\end{eqnarray}
Eq.~(\ref{eq:polf}) assumes time-reversal symmetry.  We use the tetrahedron
method for the Brillouin zone (BZ) summation in Eq.~(\ref{dieele}) following
Ref.~\onlinecite{rath75}.  We first calculate the contribution to $\Pi$
proportional to the imaginary part of the second line in
Eq.~(\ref{eq:polf}), and determine the rest of $\Pi$ by Hilbert
transformation (Kramers-Kr\"onig relation).  Such approach significantly
reduces the computational time required to calculate $\Pi$.

The correlation part of $\Sigma$ is
\begin{eqnarray}
\langle \Psi_{{\bf q}n}|\Sigma_{\rm{c}}(\omega) |\Psi_{{\bf q}n} \rangle
&& = \sum^{\rm BZ}_{\bf k}  \sum^{\rm  All}_{n'} \sum_{IJ}
\langle \Psi_{{\bf q}n}| \Psi_{{\bf q-k}n'} \tilde{M}^{\bf k}_I \rangle
\langle \tilde{M}^{\bf k}_J \Psi_{{\bf q-k}n'} | \Psi_{{\bf q}n} \rangle  \nonumber \\
&& \times \int_{-\infty}^{\infty} \frac{i d\omega'}{2 \pi}
W^{\rm{c}}_{IJ}({\bf k},\omega')
\frac{1}{-\omega'+\omega-\varepsilon_{{\bf q-k}n'} \pm i \delta}.
\label{eq:sigc}
\end{eqnarray}
Here $-i \delta$ is for occupied states; $+i \delta$ is for unoccupied
states.  $W^{\rm{c}} \equiv W - v$.

%

$GW$ calculations usually approximate Eq.~(\ref{eq:e1shot}) by
\begin{equation}
\Ekn=\ekn+Z_{{\bf k}n}[\langle\Psikn|
\Sigma^\val(\bfr,\bfr',\ekn)|\Psikn \rangle
- \langle\Psikn |V_{\rm xc}^{\rm{LDA}}
([n^\val],\bfr)|\Psikn \rangle]
\label{eq:qpgwnc}
\end{equation}
where $\Sigma^\val$ and $V\xc^{\rm{LDA}}([n^\val])$ are calculated only from eigenfunctions belonging to \val.
In the present method we calculate the $\Ekn$ including
the core contributions from
\begin{eqnarray}
\Ekn=\ekn+
Z_{{\bf k}n}\times [\langle \Psikn|
\Sigma_{\rm x}(\bfr,\bfr')+
\Sigma_{\rm c}(\bfr,\bfr',\ekn)|\Psikn \rangle \nonumber \\
- \langle\Psikn|V\xc^{\rm{LDA}}([n^{\rm total}],\bfr)|\Psikn \rangle].
\label{eq:qpgw}
\end{eqnarray}
Note $V\xc^{\rm{LDA}}([n^{\rm total}],\bfr)$ is for entire density $n^{\rm
total}$.  As $n'$ in \req{eq:sigx} can be divided between \core\ and \val,
we have $\Sigma_{\rm x} = \Sigma^\core_{\rm x} + \Sigma^\val_{\rm x}$.  In
this paper, we neglect the \core\ contribution to $\Sigma_{\rm c}$.  In
Sec.~\ref{sec:core} we examine in some detail the contributions by shallow
cores to correlation by including them in VAL using local orbitals, and
will see that their contribution is small except for very shallow cores.

In short, no important approximation is made other than the $GW$
approximation itself; and it is to the best of our knowledge the only
implementation of $GW$A that makes no significant approximations.  Results
depend slightly on what kind basis set used to generate $G$ and $W$, as we
will show, and also on the tolerances in parameters used in the
$GW$-specific part of the calculation.

LMTO-based calculations presented here employ a widely varying set of basis
functions, ranging from $\sim$20$-$90 orbitals per atom, as described in
more detail below.  They typically consist of a basis of {\em
spdfg}+{\em{}spd} orbitals centered on each atom, some floating orbitals
and sometimes local orbitals.  In the Si calculations local $p$ orbitals of
either of $2p$ character or of $4p$ character were used, as we describe
below.  For the $GW$ part of the calculation, the Si results shown below
use parameters representative of the various system studied: LMTO basis
functions are re-expanded in plane waves to a cutoff of 3.3~a.u. in the
interstitial region, i.e. $\left|\bf{k}+\bf{G}\right|<3.3$~Bohr$^{-1}$ in
the second term of Eq.~(\ref{def:psiexp}).  The IPW part of the mixed basis
used to expand $v$, $\Pi$, and $W$ used a cutoff
$\left|\bf{k}+\bf{G}\right|<3.0$~Bohr$^{-1}$; the product basis part
consisted of 90--110 Bloch functions/atom (we use different product basis
for $\Sigma^\core_{\rm x}$ and $\Sigma^\val_{\rm x}$).  Augmentation sphere
radii were chosen so that spheres approximately touched but did not
overlap, and the product basis functions entering into the mixed basis
$\tilde{M}$ in Eqns.~(\ref{eq:polf}) and (\ref{eq:sigc}) were expanded to
$l$=5. In the calculation of Eq.~(\ref{eq:sigc}) the poles of $G$ were
Gaussian broadened by $\sigma$=0.003 or 0.01~Ry. These parameters
correspond to rather conservative tolerances: tests at tighter tolerances
in these parameters change the QPE by $\sim$0.01~eV.  (Systematic checks
were performed for each material studied.)  The tetrahedron method was used
for $\Pi$ with a $6\times{}6\times{}6$ $k$-mesh (doubling the number of
points in the energy denominator) except where noted.  The same mesh is
used to calculate $\Pi$ and $\Sigma$.  This $k$ mesh is reasonably well
converged, systematically overestimating conduction-band states by
$\sim$0.02~eV in Si and similar semiconductors relative to the fully
$k$-converged result.\cite{kconvergence}

\section{Convergence in quasiparticle energies: number of unoccupied states}
\label{sec:nconvergence}

\begin{figure}[htbp]
\centering
\includegraphics[angle=0,width=.45\textwidth,clip]{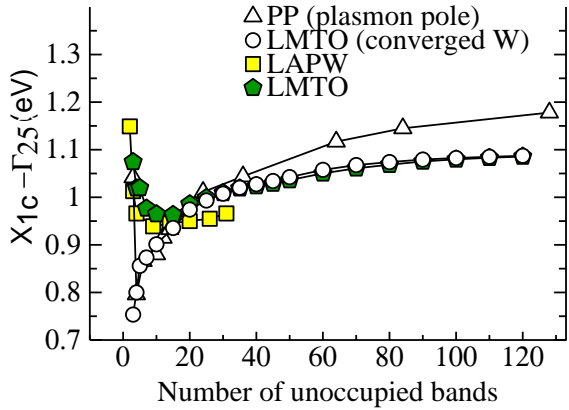}
\caption{$\Gamma_{25v}${}$\to$X$_{1c}$ gap in Si in $GW$A
    as a function of the number of unoccupied states $N'$.
{\bf Filled(Yellow) squares}: LAPW  $GW$A taken from Ku and Eguiluz\cite{weiku04}.
    The authors presented data only for $N'<31$.  Also, their data was
    given for the minimum gap, so we shifted their
    results by +0.14~eV to estimate the $\Gamma_{25v}${}$\to$X$_{1c}$ gap.
    Some checks show that the shift should be $\sim$0.14~eV, approximately
    independent of $N'$.
{\bf Filled(Green) pentagons}: LMTO results varying $N'$ both $G$ and $W$.
    LMTO results were shifted by -0.02~eV to correct for incomplete $k$
    convergence{\protect\cite{kconvergence}}. The
    Si 2$p$ was included in the valence using a local orbital
    (the LAPW calculation of Ku and Eguiluz did not).  The total dimension of the LMTO basis is 180.
    Results from the same basis are shown as Filled(green) pentagons in Fig.~\ref{fig:gapvsinversen},
     and also Filled(green) pentagons No.~(13) in Fig.~\ref{fig:gapvsbasis}.
{\bf Open triangles}:
    PP$GW$ from Ref.~\onlinecite{tiago04}, which included the Si $2p$ levels as part of the
   valence, and in which $W$ was computed using the plasmon-pole approximation.
{\bf Open circles}:
    LMTO results varying $N'$ in $G$ but not $W$.
}
\label{fig:gapconvergence}
\end{figure}

In Fig.~\ref{fig:gapconvergence}, we show the $\Gamma_{25v}${}$\to$X$_{1c}$ gap
for Si computed by Eq.~(\ref{eq:qpgw}) as a function of the number of
unoccupied states $N'$ in Eqns.~(\ref{eq:polf}) and (\ref{eq:sigc}): i.e.
summation $n'$ over unoccupied states is restricted to $n'<N'$.
Fig.~\ref{fig:gapconvergence} depicts our main with pentagons
(Si 2$p$ treated as \val).  It tracks well the all-electron results of
Ku and Eguiluz~\cite{weiku04}, which used
an LAPW method, except their data is $\sim$0.05~eV less than ours.
However, their results are limited to $N'<31$, which is not sufficient to
analyze convergence for large $N'$.  If one assumes the LAPW converges with
$N'$ at the same rate as the LMTO case, their best result with $N'=31$
should be $\sim$0.1~eV less than the converged value.  Indeed a very recent
calculation of the same system by Friedrich, Schindlmayr, Bl{\"u}gel and
Kotani~\cite{Christoph06}, based on LAPW with an LDA basis of $\sim$300
basis functions, showed $N'$-convergence similar to LMTO.

Tiago, Ismail-Beigi, and Louie\cite{tiago04} presented a PP$GW$ calculation
of some QPE in Si, Ge, and GaAs where they included the higher-lying core
states into the valence so as to assess the effect of the core.  They
monitored the rate of convergence in QPE with $N'$; their data for Si are
shown as open triangles in Fig.~\ref{fig:gapconvergence}.  There are some
similarities, but also two discrepancies:
\setcounter{Alist}{0}
\begin{list}{(\roman{Alist})\,}{\leftmargin 12pt \itemindent 22pt \usecounter{Alist}}
\item For $N' \lesssim 30$, the behavior is rather different.
\item In the asymptotic region $N' \gtrsim
30$, the PP$GW$ and LMTO results converge at somewhat different rates.
\end{list}

In order to examine point (i), we tried LMTO calculations where $W$ is
fixed (i.e. $N'$ is truncated only in Eq.~(\ref{eq:defg})).  This
calculation (open circles in Fig.~\ref{fig:gapconvergence}) tracks well the
PP$GW$ result for $N'\lesssim 30$.  This looks reasonable because the
PP$GW$ is combined with the plasmon-pole approximation, which satisfies the
sum rule for ${\rm{}Im}\,\epsilon^{-1}$ for any $N'$; thus $W$ converges
rather quickly with respect to $N'$.  However, the two LMTO calculations
show little difference in the asymptotic behavior, which means that it is
controlled by $N'$ in Eq.~(\ref{eq:sigc}), as was already discussed by
Tiago et al.

It can be seen that the $N'$ dependence of either LMTO calculation is
slightly different than the PP$GW$ result for both intermediate and large
$N'$: the change in the $\Gamma_{25v}${}$\to$X$_{1c}$ gap from $N'=35$ to
$N'=60$ for PP$GW$ is roughly twice the change obtained by the LMTO method.
As we noted in Sec.~\ref{sec:methodology}, LMTO-LDA eigenvalues are very
close to the full APW results in this energy range (see
Fig.~\ref{fig:sildabands}). This indicates that the eigenfunctions are also
precise.  Moreover, Friedrich et al.~\cite{Christoph06} compare LDA-APW
eigenvalues to an LAPW+local orbitals method; the three sets of eigenvalues
(APW and LMTO+local orbitals, LAPW+local orbitals) are very close to one
another.  By contrast, the LDA energy bands computed by either conventional
LAPW or PP methods correspond to APW eigenvalues far less well; see
Ref.~\onlinecite{Christoph06}.  Finally, the dependence on $N'$ as computed
by the LAPW+local orbitals method is essentially similar to the present
LMTO results.  When these observations are considered as a whole, they
suggest that what discrepancy does exist between LMTO+local orbitals (or
LAPW+local orbitals) and PP$GW$ may be an artifact of the pseudopotential
construction in the PP$GW$ method.  We cannot rule out possible limitations
to the present method, however.  Differences with PPGW are small in
absolute terms.  Even though eigenvalues generated by LMTO and LAPW+local
orbitals are very close to APW eigenvalues, eigenfunctions may be less well
described.  And even though the LMTO and LAPW hamiltonians are very
different, the QPEs are generated by a common $GW$ code. If there is some
limitation in the numerical procedure, it would be common to both LMTO and
LAPW calculations.

\begin{figure}[htbp]
\centering
\includegraphics[angle=0,width=.45\textwidth,clip]{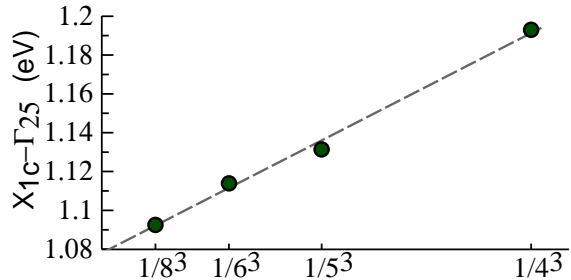}
\caption{$\Gamma_{25v}\to\rm{X}_{1c}$ gap in Si, as a function $1/n^3_k$,
  where $n^3_k$ is the number of $k$-points in the full Brillouin zone.
  The dependence on $1/n^3_k$ shown for $\Gamma_{25v}\to\rm{X}_{1c}$ is
  essentially same for all of the unoccupied QPEs we examined.  Gaps for
  $n_k$=6, $n_k$=5, and $n_k$=4, exceed the $n_k$=8 case by 0.02~eV,
  0.04~eV, and 0.10~eV, respectively.  We can also estimate what the
  $n_k\to\infty$ gap would be by extrapolating the approximately linear
  dependence on $1/n^3_k$ to zero (dashed line). The $n_k=8$ case
  apparently overestimates the converged result by 0.01~eV.  }
\label{fig:nkconvergence}
\end{figure}

It is also possible that the calculation by Tiago et al. suffers from
incomplete $k$-convergence.  Their PP$GW$ used a $4\times{}4\times{}4$
$k$-mesh.  $k$-convergence is mainly limited by divergent behavior for
$|{\bf k}| \to 0$ in Eq.~\ref{eq:sigx}.  To treat this divergence, we use
the offset-$\Gamma$ method, which was originally developed by
ourselves\cite{kotani02} and is now used by other groups
\cite{alouani03,yamasaki03}.  It is essentially equivalent to techniques
that treat the divergent part analytically, as is typically done by PP$GW$
practitioners.  Fig.~\ref{fig:nkconvergence} shows the dependence of
$\Gamma_{25v}\to\rm{X}_{1c}$ gap on $n_k$, but in general all of conduction
bands shift nearly rigidly with changes in $n_k$ ($n_k$ = number of linear
divisions of the $k$-mesh in the Brillouin zone).  Bandgaps are
approximately linear in the reciprocal of the total number of points,
$1/n^3_k$.  The figure shows that a $4\times{}4\times{}4$ $k$-mesh
overestimates the $k$-converged gap by $\sim$0.1~eV~\cite{kconvergence}.
This may explain most of the remaining discrepancy between the PP$GW$
calculations of Tiago et~al. and the present results.

In Sec.~\ref{sec:core} we analyze the dependence of QPEs on the core
treatment.  Proper treatment of the core is somewhat
subtle\cite{corelocal}, and we use the local orbitals for the analysis.
Because they are already nearly exact solutions of the LDA for the states
they constructed to represent, they minimally hybridize with other basis
functions; consequently any higher lying \core\ state can be readily be
converted into a valence state with minimal perturbation of the LDA basis.
Use of local orbitals enables us to investigate how different kinds of core
contributions affect QPEs in a well-controlled and systematic way.  We show
that differing treatments of the Si $2p$ core only slightly affect QPEs;
similar results are found for other deep cores.  A significantly larger
dependence is apparently found using the PP$GW$ method.

\section{Convergence in quasiparticle energies: basis dependence}
\label{sec:basisconvergence}


\begin{figure}[htbp]
\centering
\includegraphics[angle=0,width=.45\textwidth,clip]{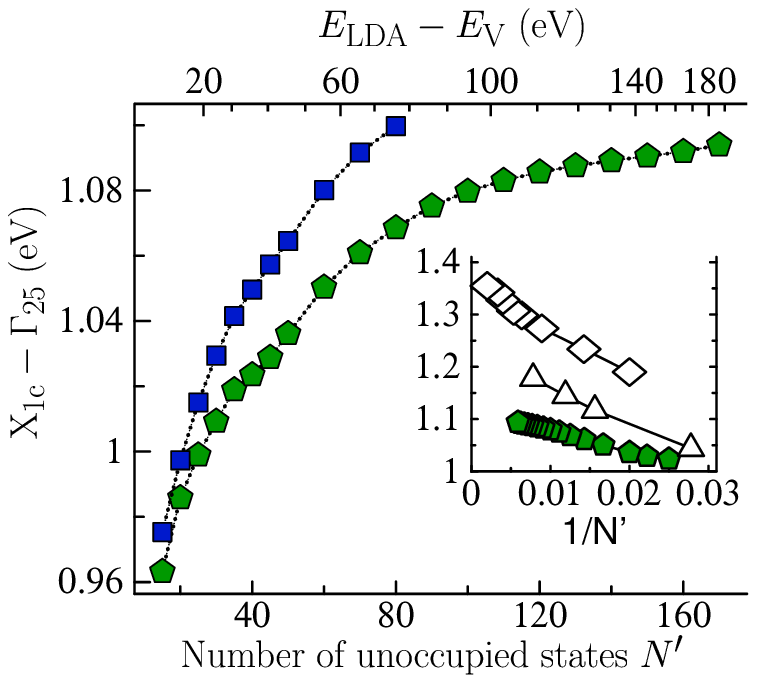}
\ 
\includegraphics[angle=0,width=.267\textwidth,clip]{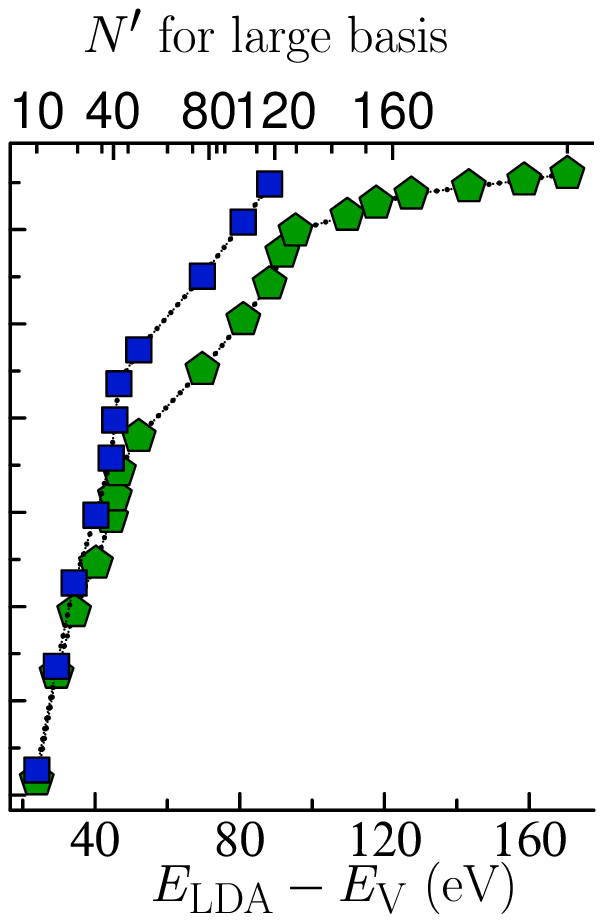}
\caption{Left panel shows $\Gamma_{25v}\to\rm{X}_{1c}$ gap in Si, as a
  function of number of unoccupied states ${N'}$ for a smaller basis
  (squares) and a larger basis (pentagons).  The latter are redrawn from
  Fig.~\ref{fig:gapconvergence}.  Top horizontal scale shows an approximate
  relation between energy and $N'$ in the large basis (interpolated from
  levels at $\Gamma$).  Right panel contains the same data but reverses the
  top and bottom horizontal scales.  (Had $N'$ in the upper horizonatal
  axis been drawn for the small basis (squares), the scale would be a
  little different: the last data point corresponds to $N'=82$ instead of
  120.)  Inset compares convergence in {X}$_{1c}$ as a function of $1/N'$
  to a PP$GW$ calculation that includes $2p$ states in the
  valence~\cite{tiago04} (triangles) and a PP$GW$ calculation that does
  not~\cite{Steinbeck00} (diamonds).  LMTO data were shifted by -0.02~eV to
  correct for incomplete $k$ convergence{\protect\cite{kconvergence}}.  The
  small-basis data (squares) in the right panel were further shifted by
  -0.01~eV\cite{numprecision} to clarify how large- and small-basis data
  diverge as the energy increases.  }
\label{fig:gapvsinversen}
\end{figure}

\begin{figure}[htbp]
\centering
\includegraphics[angle=0,width=.45\textwidth,clip]{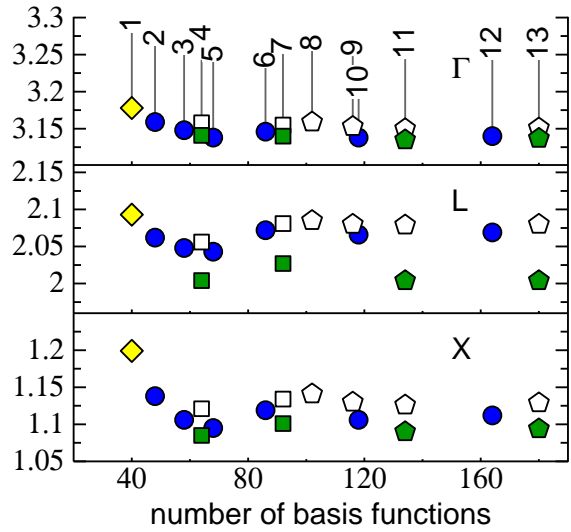}
\caption{QPE in the $GW$A at $\rm{\Gamma}_{15c}$ (top),
 $\rm{L}_{1c}$ (middle), and $\rm{X}_{1c}$ (bottom) in Si relative
 to the valence band maximum, using different basis sets in the present FP-LMTO.
 Abscissa is the total number of basis functions $N$.
 Yellow diamonds are a minimal basis (see text).
 All results depicted by fill circles contain no local orbitals.  Those
 depicted by filled (empty) squares or pentagons include the Si $2p$ (Si $3p$)
 as local orbitals.  See text for further description.
}
\label{fig:gapvsbasis}
\end{figure}

Here we study the convergence in QPEs as the LMTO basis set changes,
retaining all the eigenfunctions for a given basis in the calculation ($N'$
encompasses all unoccupied states).  A given LMTO basis defines a finite
Hilbert space of eigenfunctions; the $GW$A is a well-defined procedure in
that space, and we can study how the QPEs change as the Hilbert space is
refined.
This procedure corresponds more closely to analyses of basis set
convergence common in other kinds of calculations (e.g. LDA and
Hartree-Fock).  We can also anticipate that it will be smoother than the
$N'$ truncation of Sec~\ref{sec:nconvergence}; indeed this will turn out to
be the case (see especially Fig.~\ref{fig:gapvsbasis}): the band gaps are
insensitive to the choice of basis once a certain level of completeness is
reached.  It its also obviously true that the Hilbert space depends on the
choice of basis constructing it.  Therefore, the results presented here are
specific to the LMTO basis described in Sec.~\ref{sec:methodology}, and in
particular \emph{what kinds} of orbitals are included, e.g. orbitals of $f$
or $g$ character, or local orbitals that correct the linearization inherent
in most of the standard methods (LMTO, LAPW, and the construction of a
norm-conserving PP). By adding different kinds of orbitals we can identify
how different parts of the Hilbert space, (most notably corrections to
linearization common to most methods) affect QPEs.  Since the LMTO basis is
tailored to the crystal potential, the LDA eigenfunctions converge more
rapidly with the basis dimension than do plane-wave based basis
sets\cite{lmtonote}. Consequently we might expect a more rapid convergence
in the $GW$A QPEs.  On the other hand, by transformation to, e.g., Wannier
functions, it should be possible to design a generic scheme that exhibits
similarly rapid convergence.

Initially, we compare in Fig.~\ref{fig:gapvsinversen} the dependence of the
$\Gamma_{25v}${}$\to$X$_{1c}$ gap in Si on $N'$ for two basis sets: one
relatively small and another relatively large.  The right panel of
Fig.~\ref{fig:gapvsinversen} compares the same data, but the horizontal
scale corresponds the energy of state $N'$ at $\Gamma$.  Also in this
panel, the small basis-data at small energy was artificially shifted down
by 0.01~eV to make it easier to see how the two cases depend on
energy.\cite{numprecision}.  The difference, initially 0.01~eV, increases
by an additional 0.01~eV as the energy $E$ approaches $\sim{}E_f+50$~eV.
For higher energies, the discrepancy between the two increases more
rapidly.  This is because the ability of the small LMTO basis to describe
eigenvalues above $\sim{}E_f+40$~eV begins to degrade.

However, we can see that the gaps at the respective maximal $N'$ (i.e. all
unoccupied states included) are in good agreement.  Including all
unoccupied states in a limited basis is a another kind of hilbert space
truncation, but it is also well defined: use use the Hilbert space of
eigenfunctions basis consisting of LMTO eigenfunctions and their products
(Sec.~\ref{sec:methodology}), and apply the $GW$A within that Hilbert
space.  The LMTO basis can efficiently choose the important part of the
Hilbert space tailored to the crystal potential.  Thus good agreement need
not be some fortutitous artifact of this particular pair of the LMTO basis
sets, even though the maximal $N'$ is small in light of a traditional $N'$
cutoff analysis\cite{ntrunc}.  Indeed the $N'$ cutoff of
Sec~\ref{sec:nconvergence} may choose the Hilbert space less well,
especially since that kind of truncation is not smooth.  Below we present
detailed analysis of the dependence on the basis set to justify the good
agreement in Fig.~\ref{fig:gapvsinversen}\cite{numprecision}: the band gaps
are insensitive to the choice of basis once a certain level of completeness
is reached.

Fig.~\ref{fig:gapvsbasis} shows results of a systematic study of the
convergence in the first unoccupied QPE at $\Gamma$, X, and L in Si with
progressively larger basis sets.  LDA eigenvalues are not shown because
they are the same within $\sim 0.01$~eV for all cases (0.60~eV for X,
1.42~eV for L, 2.52~eV for $\Gamma$).
These data comprise very diverse basis sets, particularly for the LMTO
method which traditionally uses a minimal basis.  Some details concerning
these sets help explain in what manner convergence is reached:
\setcounter{Alist}{0}
\begin{list}{\S \arabic{Alist}\,}{\leftmargin 12pt \itemindent 22pt \usecounter{Alist}}

\item {\bf Filled(yellow) diamonds (1)} includes \emph{spdfsp} atom-centered functions,
      and is the only basis without floating orbitals.  There are no local
      orbitals; Si 1$s$,2$s$,2$p$ are \core.

\item {\bf Filled(blue) circles (2) and (3)} add floating orbitals of $sp$
      and of $spd$ character, respectively.  Their effect is to cause QPEs
      to {\em decrease} slightly relative to (1).  Adding still more
      floating orbitals (even large numbers of them) shift QPEs by
      $\sim$0.01~eV.

\item {\bf Other Filled(blue) circles (5, 6, 10, 12)} include still more
      envelope functions comprised of a mixture of atom-centered
      functions and floating orbitals, but adding no local orbitals.

\item {\bf Filled(green) Square and Open Square (4, 7)}
      correspond to (3, 6), respectively, but adding a local orbital
      (Green: Si 2$p$, Open: Si 4$p$).
      When the $2p$ is included as \val, \core\ consists of Si1$s$,2$s$ only.
      A local orbital of \emph{either} Si2$p$ or Si4$p$ shifts QPEs---in
      roughly equal but opposite directions.

\item {\bf Filled(green) Pentagon and Open Pentagon, (8,9,11,13)}
      include an additional Si4$d$ local orbital.
      (11) corresponds to (10)+(Si2$p$ or Si4$p$)+ Si4$d$.
      (13) corresponds to (12)+... as well.
      (8) is (6)+Si4$p$+Si4$d$. The effect of Si4$d$ is small.
\end{list}

These points show in a compelling way that once the basis reaches a certain
level of completeness, the change of QPE with further enlargement is very
small.  Set (1), which consists only of atom-centered functions, is
somewhat incomplete except inside the augmentation spheres where the
eigenfunctions are constructed out of linear combinations of
$\{\phi,\dot\phi\}$.  Considering the open structure of zincblende, such a
basis may be expected to be less complete in the interstitial.  Comparison
basis sets without local orbitals (circles) with set (1) shows that this
particular purely atom-centered basis is slightly deficient for reliable
calculation of QPE, since the addition of floating orbitals induces a
($k$-dependent) reduction in the conduction band of $\sim{}0.02-0.10$~eV.
It is an open question whether a still more sophisticated atom-centered
basis\cite{nmto} would be adequate to describe the interstitial.

Once the interstitial is reasonably complete ($cf.$ sets (3) and higher),
there is an almost negligible dependence on basis \emph{provided} no
orbitals are included that extend the linear method or alter how the core
is treated.  Basis sets marked by a common symbol (squares, circles,
pentagons) share essentially the same Hilbert space in the
\emph{augmentation} regions; only the basis set corresponding to the
\emph{interstitial} region changes.  The variation is $\pm$0.01~eV for a
wide range of basis sets\cite{numprecision}.

Fig.~\ref{fig:gapvsbasis} also gives us some insight to the limitations of
the linear method.  Basis sets (3) and (4), which differ only in how the Si
$p$ channel is treated inside the augmentation region, affect QPEs more
strongly than radically enlarging the Hilbert space of the envelope
functions---compare (3)$\to$(4) and (3)$\to$(12).  Envelope functions
affect only the interstitial; they negligibly affect the Hilbert space of
the augmentation region.  For the latter it is largely irrelevant how many
envelope functions are used---and consequently the size of $N'$ entering
into Eqns.~(\ref{eq:polf}) and (\ref{eq:sigc}). What \emph{is} relevant is
the completeness of $\{\phi,\dot\phi\}$, and results are independent of
basis dimension provided that \emph{the entire} $\{\phi,\dot\phi\}$
\emph{Hilbert space is included}.  Said another way, the LMTO method is by
design reasonably complete over a certain energy window in the augmentation
spheres, more or less independent of the envelope functions.  A similar
story may be told for the interstitial: sets (3, 6, 10, 12) differ in the
number of envelope functions by as much as a factor of three, but QPEs are
unchanged within $\pm$0.01~eV.  QPEs shift when
$\{\phi,\dot\phi\}${}$\to${}$\{\phi,\dot\phi,\phi^{4p}\}$ or
$\{\phi,\dot\phi\}${}$\to${}$\{\phi,\dot\phi,\phi^{2p}\}$, essentially
independently of the number of envelope functions (3$\to$4, 6$\to$7,
10$\to$11, 12$\to$13).

\begin{table}
\caption{
\baselineskip 12pt
QPEs of the first unoccupied state at $\Gamma$, L, and X, for different
basis sets, in eV (relative to valence maximum).  Columns $n_a$, $n_f$, and
$n_l$ denote the number of atom-centered functions, the number of floating
orbitals, and the number of local orbitals respectively.  The hamiltonian
dimension is the sum of these numbers.  Experimental data are adjusted for
spin-orbit coupling by adding 1/3 of the splitting in the $\Gamma_{15}$
valence bands.  The first four CdO basis sets are identical to the last
four except for the addition of local orbitals in the O 3$p$ and Cd $5d$
channels.  A $6\times{}6\times{}6$ $k$-mesh was used in these
calculations.}
\label{tab:gapvsbasis}
\begin{tabular}{l@{\hspace*{2em}}|*3{@{\quad}r}@{\hspace*{2em}}|*3{@{\quad}r}}
Data type     & $n_a$  & $n_f$ & $n_l$   &$\Gamma$  &L     &  X \cr
\colrule
\vbox{\vskip 2pt}
& \multispan5 CdO          \cr
Expt+0        &        &       &         &  0.84    &       &    \\
LDA           &  59    &  18   &  12     & -0.53    &  4.26 & 3.58 \\
$GW$          &  59    &  18   &  12     &  0.14    &  5.18 & 4.97 \\
              &  59    &  50   &  12     &  0.10    &  5.14 & 4.92 \\
              &  82    &  66   &  12     &  0.10    &  5.16 & 4.89 \\
              &  82    &  82   &  12     &  0.10    &  5.16 & 4.88 \\
              &  59    &  18   &   3     & -0.01    &  5.05 & 4.78 \\
              &  59    &  50   &   3     & -0.06    &  5.01 & 4.73 \\
              &  82    &  66   &   3     & -0.02    &  5.05 & 4.73 \\
              &  82    &  82   &   3     & -0.02    &  5.06 & 4.74 \\
\colrule
\vbox{\vskip 2pt}
& \multispan5 Ge          \cr
Expt+0.10     &        &       &         &  1.00    &  0.88 & 1.20 \\
LDA           &  50    &  18   &  10     & -0.12    &  0.07 & 0.65 \\
$GW$          &  50    &  18   &  10     &  0.80    &  0.65 & 0.94 \\
              &  68    &  18   &  10     &  0.84    &  0.68 & 0.97 \\
              &  82    &  50   &  10     &  0.83    &  0.67 & 0.96 \\
              &  82    &  82   &  10     &  0.82    &  0.67 & 0.96 \\
\colrule
\vbox{\vskip 2pt}
& \multispan5 GaAs        \cr
Expt+0.11     &        &       &         &  1.63    &  1.96 & 2.11 \\
LDA           &  42    &  18   &   6     &  0.34    &  0.86 & 1.34 \\
$GW$          &  42    &  18   &   6     &  1.44    &  1.68 & 1.79 \\
              &  68    &  18   &   6     &  1.46    &  1.69 & 1.79 \\
              &  82    &  50   &   6     &  1.44    &  1.66 & 1.77 \\
              &  82    &  82   &   6     &  1.43    &  1.66 & 1.77 \\
              &  82    &  82   &  11     &  1.43    &  1.68 & 1.81 \\
\colrule
\end{tabular}
\end{table}

Table \ref{tab:gapvsbasis} shows three other materials
(CdO, Ge, and GaAs). We can see
(1) rapid convergence in QPEs as the basis is enlarged
for a \emph{fixed} set of augmentation functions; and
(2) extensions to a
linear augmentation affect QPEs in an manner approximately
independent of the total dimension of the Hamiltonian.  (In GaAs,
both $3d$ and $4d$ must be included as \val. If not, significant errors result\cite{kotani02}).
We have
tested a number of other materials as well, and these trends appear to be
rather general.  As might be expected, the number of basis functions needed
to make the Hilbert space reasonably complete depends somewhat on the
elements involved.  The heavier atoms have larger radii and consequently
slower $l$-convergence in the number of envelope functions needed; also $d$
orbitals often play an important role.  More orbitals are required to make
the basis complete when heavier atoms are involved.

As noted, the linear $\{\phi,\dot\phi\}$ Hilbert space is already
reasonably complete in the case of Si.  But this is not true in general:
oxides and nitrides form a materials class where the effect is
significantly larger.  CdO is one such example (CdO forms in the NaCl
structure; the valence band maximum falls at L and the conduction band
minimum falls at $\Gamma$.)  As happens for Si, there is a weak dependence
on basis when the number of envelope functions is changed and the Hilbert
space of the augmentation is held constant.  But the QPEs change by
$\sim 0.15\pm0.05$~eV when the O 3$p$ and Cd $5d$ states are added, as
Table~\ref{tab:gapvsbasis} shows.  (In this particular case it is the O
$3p$ contribution that is dominant; however, cases arise when the
contributions from high-lying $d$ or $f$ orbitals can be of order $1-2$~eV.
NiO is such a case\cite{faleev04}.)

The inset of Fig.~\ref{fig:gapvsinversen} shows some PP$GW$ results for
reference.  Based on the observation that cutoff in the Hilbert space
should be important, PP$GW$ data by Tiago et al. should be extrapolated to
$1/{N'}\to{}0$ because they used a very large LDA basis.

\section{Core contributions to $\Sigma_{\rm{c}}$}
\label{sec:core}

In a series of papers, Shirley and co-workers analyzed the effects of the
core on QPE in atoms (Shirley, Mit{\'a}s and Martin,
Ref.~\onlinecite{Shirley91}, Shirley and Martin,
Ref.~\onlinecite{Shirley93}) semiconductors (Shirley, Zhu and
S. G. Louie~\cite{Shirley92,Shirley97}) within the pseudopotential
framework.  Approximate core contributions to both Eq.~(\ref{eq:polf}) and
(\ref{eq:sigc}) were evaluated.
They also compared pseudopotentials constructed from both LDA
exchange and from Hartree-Fock (HF) exchange for atoms and
molecules\cite{Shirley90}, and incorporated pseudopotentials of both types
in studying core effects\cite{Shirley92,Shirley97}.
They found sizable shifts in QPE in Si, and rather dramatic and
$k$-dependent shifts in Ge and GaAs.  These analyses highlight the
importance of core effects.
However, the decomposition of the various core contributions in
Ref.~\onlinecite{Shirley97} is somewhat involved, and it is rather closely
tied to the pseudopotential construction that was a part of their
implementation.  This makes it a little difficult to disentangle the
various contributions.


Here we examine contributions from the shallowest cores to
$\Sigma_{\rm{c}}$ within the framework of our $GW$.  As we noted, all the
eigenfunctions are divided into two groups \core\ and \val\ as explained in
Sec.~\ref{subsec:allgw}.  Using local orbitals we can represent the
shallowest cores in \val.  To distinguish true core effects from artifacts
of implementation\cite{corelocal}, we include these cores in the valence
with local orbitals and treat them in a special way, as described below.
We denote such eigenfunctions as \itcore, and the rest as \itval.  Thus we
distinguish three kinds of orbitals: CORE, \itcore, and \itval:
\begin{eqnarray}
\quad \pstree[nodesep=2pt,levelsep=20pt]{\TR{\hbox{All eigenfunctions}}}
  {\pstree{\TR{\core}}{}
   \pstree{\TR{\val}}
     {\pstree{\TR{\itcore}}{}
      \pstree{\TR{\itval}}{}
  }
}
\label{eq:}
\end{eqnarray}
In Si, for example, we use: \core=$\{1s,2s\}$, \itcore=$\{2p\}$ and
\itval=$\{3s,3p,3d,...\}$.  Because the \itcore\ states are well separated
from higher lying states, $G$ can be partitioned into
$G=G^\core+G^\itcore+G^\itval$.  $\Sigma_{\rm{x}}$ is always calculated from
the entire $G$, while $\Sigma_{\rm{c}}$ is calculated from \itcore\ and
\itval\ only (we do not consider any case where some portion of the self-energy
is supplied by the LDA):  $\Sigma_{\rm{c}} = i (G^\itcore + G^\itval) W^{\rm c}$, where
$W^{\rm c}=W[\Pi]-v$, and $\Pi$ is calculated from $G^\itcore + G^\itval$.  Thus \itcore\
states contribute to $\Sigma_{\rm{c}}$ directly through $G$ in $i G W^{\rm c}$,
and also through $W^{\rm c}$.  We resolve these contributions; that is, we
calculate $\Sigma_{\rm{c}}$ in one of four ways:
\begin{list}{({\it \roman{Alist}}\,)}
{\leftmargin 12pt \itemindent 22pt \usecounter{Alist}\addtocounter{Alist}{0}}
\item neglect the \itcore\ contribution to $\Sigma_{\rm{c}}$ entirely: i.e.
      $\Sigma_{\rm{c}} = i G^\itval (W[\Pi^{\itval}]-v)$,
      where $W[\Pi^{\itval}]$ means that $\Pi$ is calculated from $G^\itval$ only.
      We denote this as ``exchange-only \itcore''.

\item neglect the \itcore\ contribution to screening :
      $\Sigma_{\rm{c}} = i (G^\itcore + G^\itval) (W[\Pi^{\itval}]-v)$.

\item neglect the \itcore\ contribution to $G$:
      $\Sigma_{\rm{c}} = i G^\itval (W[\Pi]-v)$.

\item $\Sigma_{\rm{c}} = i (G^\itcore + G^\itval) (W[\Pi]-v)$:
       there is no distinction between \itcore\ and \itval \ states.

\end{list}

\begin{table}
\caption{QPE of the first unoccupied state at $\Gamma$, L, and X relative
 to top of valence, for core treatments $(i)-(iv)$ as described in the
 text, in eV.  States and corresponding eigenvalues
 $\varepsilon_c^{\rm{LDA}}$ treated as \itcore\ are: $2p$ in Si, $3d$ in Ga and
 Ge, and $2s$ in Mg.
 Si data corresponds to basis set (13) in Fig.~\ref{fig:gapvsbasis}; GaAs
 data corresponds to the 68+18+6--orbital basis in
 Table~\ref{tab:gapvsbasis}; Ge data corresponds to the 68+18+10--orbital
 basis in that table.  Here $G$ means $(G^\itcore + G^\itval)$, $W$ means
 $W[\Pi]$ (see text).  A $6\times{}6\times{}6$ $k$-mesh was used in these
 calculations.  For results with better $k$-convergence and larger basis
 sets, see Table~\ref{tab:gladgaps}.}
\label{tab:corecontr}
\begin{tabular}{l@{\hspace*{1.5em}}|@{\hspace*{1.5em}}r@{\hspace*{1.5em}}rrr@{\hspace*{1.5em}}rrr@{\hspace*{1.5em}}rrr@{\hspace*{1.5em}}rrr}
              & $\varepsilon_c^{\rm{LDA}}$
              & \multispan3 \hfil$(i):\,G^\itval,W[\Pi^\itval]$\hfil& \multispan3 \hfil$(ii):\,G,W[\Pi^\itval]$\hfil&
               \multispan3 $(iii):\,G^\itval{},W$\hfil   & \multispan3 $(iv):\,G,W$ \hfil \cr
              &   & $\Gamma$ & L & X       & $\Gamma$ & L & X        & $\Gamma$ & L & X     & $\Gamma$ & L & X  \cr
\colrule
Si            & -89.6 & 3.17& 2.09& 1.14   & 3.17& 2.09&  1.14       & 3.17& 2.06& 1.15     & 3.16& 2.02& 1.11  \cr
Ge            & -24.7 & 0.98& 0.74& 0.98   & 0.96& 0.73&  0.95       & 0.88& 0.71& 0.99     & 0.84& 0.68& 0.97  \cr
GaAs          & -14.8 & 1.65& 1.83& 1.86   & 1.63& 1.82&  1.85       & 1.54& 1.75& 1.83     & 1.46& 1.69& 1.79  \cr
MgO           & -71.4 & 7.31&10.55&11.62   & 7.36&10.56& 11.62       & 7.30&10.55&11.60     & 7.36&10.55&11.60  \cr
\colrule
\end{tabular}
\end{table}

Table \ref{tab:corecontr} shows that the difference between exchange-only $(i)$
and $GW$ $(iv)$ approximations to core treatment is small in Si
($\sim$0.03~eV for $\rm{X}_{1c}$).  As expected, the adequacy of an
exchange-only core depends on how deep the core is.
The exchange-only approximation for shallow cores, such as
the Ga $3d$ and In $4d$, and the highest lying $p$ core in the column I
(Na, K, Rb) and column IIA alkali metals (Mg, Ca, and Sr), is
rather crude.  It is interesting that the core contributions to
$\Pi$ and to $G$ are \emph{not} always additive.


The difference between $(iii)$ and  $(iv)$ is in general rather
small; that is, inclusion of \itcore\ contributions of $\Pi$ alone is
sufficient to bring QPE results within 0.05~eV of the full results in
Table~\ref{tab:corecontr} except for the very shallow Ga 3$d$ channel.  For
moderately deep cores, exchange-only treatment $(i)$ is generally
adequate, as Aryasetiawan suggested.  A rough rule of thumb seems to
be: for cores whose total charge $Q_{\rm{spill}}$ outside the augmentation
radius is less than 0.01 electrons, exchange-only treatment of them results
in errors $\sim$0.1~eV or less for the lowest excited states. (This radius
may be taken as approximately half the nearest-neighbor bond length).

Inclusion of core contributions to $\Pi$ can significantly
increase the computational cost (in the Si case, leaving out the 2$p$
contribution to $\Pi$ reduces the computational cost by $\sim$40\%).
The relative smallness of corrections to exchange-only treatment, and the
observation that core contributions to $\Pi$ alone are adequate for all but
the most shallow cores, suggests that a simple approximate inclusion of
core contributions to $\Pi$, Eq.~(\ref{eq:polf}), should be adequate for
all but the most shallow cores such as the Ga $3d$. (Fleszar and Hanke
proposed a construction for pseudopotentials when core states are not
pseudized\cite{Fleszar05}.)  Supposing the core were confined to the
augmentation sphere at site $R$, we can eliminate all contributions to the
matrix element $\langle \tilde{M}^{\bf q}_I \Psi_{{\bf k}n} |\Psi_{{\bf
q+k}n'}\rangle$ except from the product-basis contribution at $R$.  Since
also the augmented part of $\Psi$ depends rather weakly on
$\dot\varphi_{Rl}$ we can neglect the $\dot\varphi_{Rl}$ contribution to
the eigenfunctions and assume that $\langle \tilde{M}^{\bf q}_I \Psi_{{\bf
k}n} |\Psi_{{\bf q+k}n'}\rangle$ only depends $n$, $n'$, $\bfk$, or
$\bfk+\bfq$ through the coefficients,
$(\alpha^{{\bfk}n}_{Ru})^*\alpha^{{\bfk+\bfq}n'}_{Ru}$.  Moreover, the core
level energy is large and negative, and nearly independent of $\bfk$ or
$n$.  Since the dominant contributions to $\Pi$ will come from coupling to
low-lying states, we can approximate $\varepsilon_{kn}-\varepsilon_{k'n'}$
by a constant, e.g.  $\ef-\varepsilon_{\rm{core}}$.  These
approximations are all modest but can vastly simplify the computation of
$\Pi^\core$.

The fact that the core spills out slightly from the augmentation region
needs to be taken into account\cite{coretrunc}.  This can readily be
accomplished by integrating the core and corresponding valence
$\varphi_{l}$ to a larger radius, and orthogonalizing $\varphi_{l}$ to the
core.  Checks show that the adjustment to $\phi_l$ is small unless the core
is very shallow, in which case the core should be treated as a valence
state.

\subsection*{Comparison to PP$GW$}
\label{subsec:ppgw}

When the highest cores are put explicitly into the valence as Tiago
et al. did, there is reasonable agreement between PP$GW$ and our results
for $sp$ semiconductors.  Comparison with the paper of Tiago et al. to
Table~\ref{tab:corecontr} below shows that there is agreement at the
$\sim$0.1~eV level in Si\cite{gapdiscrepancy} and similar agreement is
found for GaAs and Ge, with the PP$GW$ results systematically higher than
our results by $\sim$0$-$0.1~eV.  Similarly, Fleszar and
Hanke\cite{Fleszar05} calculated QPEs in the $GW$A for a variety of II-VI
semiconductors, including the highest $s$ and $p$ cores in the valence.
Their values are also in reasonable agreement with results presented in
Table~\ref{tab:gladgaps} (below), though the PP$GW$ data are systematically
higher by $\sim$0.0-0.2~eV.  (Part of the discrepancy can be traced to
contributions from high-lying $d$ states, which are included in the present
calculation using local orbitals.)  Even when the high-lying $s$ and $p$
core states are included explicitly in the valence, it still seems to be
the case that PP$GW$ band gaps are systematically
slightly larger (by $\sim$0.1~eV) in semiconductors than our $GW$ predict.

Materials involving transitions metals are rather more complicated.  In a
recent PP$GW$ calculation, Marini, Onida and del Sole analyzed the QP
valence bands of Cu\cite{Marini02}, comparing in some detail the occupied
$d$ bands to photoemission experiments.  The LDA places the position of
these levels approximately 0.5~eV closer to $\ef$ than the experiments
show.  The authors find that the $d$ bands narrow and shift downward by
approximately 0.5~eV, bringing the PP$GW$ $d$ bands into excellent
agreement with photoemission experiments.  They report that the PP$GW$
results depend rather dramatically on the treatment of the Cu core $3s$ and
$3p$ levels: that it is necessary to include both states explicitly in the
valence to obtain reasonable results.  They found that the correlation
contribution $\Sigma_c^{\rm{core}}$ from these states shifts the $d$ bands
downward $\sim$0.5~eV.

We conducted a similar calculation using the present all-electron $GW$,
and find a very different result.  In our case, the
$GW$ correction to the LDA $d$ bands is small---between 0 and 0.1~eV.
Moreover, QPEs are essentially independent of how the Cu $3p$
state is treated: the $3d$ levels change by less than 0.05~eV when the Cu
$3p$ state is explicitly included in the valence (using a $3p$ local
orbital), as compared to being treated as core at the exchange-only level.
The Cu $3p$ state is rather deep, and the weak dependence on correlation
contributions from it is consistent with the rule of thumb indicated above:
$Q_{\rm{spill}}\approx{}0.005$~electrons;
$\epsilon^{\rm{LDA}}_{3p}\approx-70$~eV.  In the Cu case, it appears
likely that the main discrepancy between PP$GW$ and our $GW$
(whether $d$ bands shift by 0.5~eV or not) originates in the discrepancies
in $\Sigma_c^{\rm{core}}$.


\section{Adequacy of $GW$A applied to a range of materials}
\label{sec:gldawlda}

In Ref.~\onlinecite{delaney04} Delaney et al. argued that $GW$A based on the LDA
eigenfunctions and eigenvalues, is an adequate (or better) approximation
than self-consistent $GW$.  It is apparently the case that self-consistency
worsens agreement with experiment for the Be atom.  Moreover, Holm and von
Barth~\cite{holm98} found that the valence bandwidth of the homogeneous
electron gas is considerably worsened by self-consistency; similarly a
self-consistent $GW$ calculation Ku and Eguiluz resulted in an overestimate
of the valence bandwidth in Ge\cite{weiku02}.  Thus, self-consistency of this
type has shortcomings.  On the other hand, even in simple materials such as
$sp$ semiconductors, $GW$A bandgaps based on LDA eigenvalues and
eigenfunctions are {\em always} underestimated when properly calculated
\cite{kotani02,Fleszar05}.  The $GW$A based on LDA is evaluated as a
perturbation relative to LDA; thus the band gap can be poor if the LDA itself
is poor.  Thus, some kind of self-consistency is necessary to reduce the
dependence on the starting point.


In this section, we consider three points about the $GW$A based on LDA,
Eq.~(\ref{eq:e1shot}):
\setcounter{Alist}{0}
\begin{list}{({\Alph{Alist}})\,}{\leftmargin 12pt \itemindent 24pt \usecounter{Alist}\addtocounter{Alist}{0}}
\item {\bf Use of the $Z$ factor}.   We show that using $Z$=1 in
      Eq.~(\ref{eq:e1shot}) is a way to include partial self-consistency,
      and it should be a better approximation than including the $Z$ factor.
\item {\bf Off-diagonal $\Sigma$}.  Eq.~(\ref{eq:e1shot}) is a perturbation
      treatment that involves only the diagonal matrix element of $\Sigma$.
      We consider the effect of the full $\Sigma$ in a variety of systems
      analyzing how the adequacy of $GW$A is dependent on the adequacy of
      LDA.  Even $GW$A with $Z$=1 fails for cases when the starting LDA is poor.
\item {\bf Band-disentanglement problem.}  Even when LDA eigenfunctions are reasonable,
      if eigenvalues are wrongly ordered the perturbation
      treatment can have important adverse consequences.
\end{list}

\subsection{$Z$ factor}
\label{subsec:zfactor}


Let us consider a partial kind self-consistency where only eigenvalues are
updated: both eigenfunctions and $W$ are unchanged from the LDA.  This is a
little different from the customary usual eigenvalue-only self-consistency,
where eigenfunctions are frozen but $W$ is updated.  Updating eigenvalues
widens semiconductor bandgaps.  This reduces the screening, which causes
$W$ to increase, which in turn cause gaps to increase still more.  Thus we
expect that results from such the kind of partial self-consistency we are
considering here should fall somewhere between the usual one-shot $GW$ and
the usual eigenvalue-only self-consistency.  Partial self-consistency,
while incomplete, should result in better QPEs than the standard 1-shot
$GW$, since eigenvalues shift in the right direction.  Appendix
\ref{appendix:z1} evaluates how Eq.~(\ref{eq:e1shot}) gets modified for a
model two-level system.  The result is that this kind of self-consistency
can be approximately realized by putting $Z_{{\bf k}n}=1$ in
Eq.~(\ref{eq:e1shot}).  A different justification for omitting the $Z$
factor emerged from a paper of Niquet and Gonze\cite{Niquet04}, who
calculated the interacting bandgap energy (within RPA) to obtain a
correction to the Kohn-Sham gap.  They found that the difference
is essentially Eq.~(\ref{eq:e1shot}) with $Z$=1.  Finally, a further
justification for using $Z$=1 is discussed in chapter 7 of
Ref.~\onlinecite{mahan90}.  $Z$=1 corresponds to the Rayleigh-Schr\"odinger
perturbation, $Z$ with Eq.~(\ref{zfactor}) to the Brillouin-Wigner
perturbation.  It shows the $Z$=1 scheme should be better for the
Fr\"olich Hamiltonian Mahan analyzed.

\def\mn{\hskip -04pt-}
\def\ftn[#1]{\footnotemark[#1]}
\begin{table}
\caption{
  \baselineskip 10pt
  Fundamental gap, in eV. (For Gd, QPE correspond to the
  position of the majority and minority $f$ levels relative to $\ef$; for
  Cu QPE corresponds to the $d$ level.)
  Low temperature experimental data were used when available.  QPEs in the ``$GW$'' column
  are calculated with usual $GW$A Eq.~(\ref{eq:e1shot}) and Eq.~(\ref{eq:defzfac}).
  In the ``$Z$=1'' column the $Z$ factor is taken to be unity.
  In the ``$\Sigma_{nn'}$'' column the off-diagonal parts of $\Sigma$ are included
  in addition to taking $Z$=1.
  $k$-meshes of $8\times{}8\times{}8$ $k$ and
  $6\times{}6\times{}6$ were used for cubic and hexagonal structures,
  respectively (symbol $w$ indicates the wurtzite structure).  $GW$
  calculations leave out spin-orbit coupling and zero-point motion effects.
  The former is determined from $\Delta/3$, where $\Delta$ is the spin
  splitting of the $\Gamma_{15v}$ level (in the zincblende structure); it
  is shown in the ``$\Delta/3$'' column.  Contributions to zero-point
  motion are estimated from Table 2 in Ref.~\onlinecite{Manjon04} and are
  shown in the ``ZP'' column.  The ``adjusted'' gap adds these columns to
  the true gap, and is the appropriate quantity to compare to $GW$.}
\label{tab:gladgaps}
\begin{tabular}{l|@{\hspace*{1.0em}}lcll@{\hspace*{0.5em}}c@{\hspace*{0.5em}}l@{\hspace*{0.5em}}l@{\hspace*{0.5em}}c}
                  &    LDA           &$GW$         &$GW$      & $GW$  &   Expt & $\Delta/3$   & ZP     & Adj\\
                  &                  &             &$Z$=1     & $\Sigma_{nn'}$ \\
\colrule
C                 &     4.09         & 5.48        & 5.74     & 5.77  &  5.49  & 0            & 0.34         & 5.83   \\
Si                &     0.46         & 0.95\ftn[1] & 1.10     & 1.09  &  1.17  & 0.01         & 0.06         & 1.24   \\
Ge                &  \mn0.13         & 0.66        & 0.83     & 0.83  &  0.78  & 0.10         & 0.05         & 0.93   \\
GaAs              &     0.34         & 1.40        & 1.70     & 1.66  &  1.52  & 0.11         & 0.10         & 1.73   \\
wAlN              &     4.20         & 5.83        & 6.24     &       &  6.28  & 0\ftn[2]     & 0.20         & 6.48   \\
wGaN              &     1.88         & 3.15        & 3.47     & 3.45  &  3.49  & 0\ftn[2]     & 0.20         & 3.69   \\
wInN              &  \mn0.24         & 0.20\ftn[1] & 0.33     &       &  0.69  & 0\ftn[2]     & 0.16         & 0.85   \\
wZnO              &     0.71         & 2.51        & 3.07     & 2.94  &  3.44  & 0\ftn[2]     & 0.16         & 3.60   \\
ZnS               &     1.86         & 3.21        & 3.57     & 3.51  &  3.78  & 0.03         & 0.10         & 3.91   \\
ZnSe              &     1.05         & 2.25        & 2.53     & 2.55  &  2.82  & 0.13         & 0.09         & 3.04   \\
ZnTe              &     1.03         & 2.23        & 2.55     &       &  2.39  & 0.30         & 0.08         & 2.77   \\
CuBr              &     0.29         & 1.56        & 1.98     & 1.96  &  3.1   & 0.04         & 0.09\ftn[3]  & 3.23   \\
CdO               &  \mn0.56         & 0.10        & 0.22     & 0.15  &  0.84  & 0.01\ftn[2]  & 0.05         & 0.90   \\
CaO               &     3.49         & 6.02        & 6.62     & 6.50  &$\sim$7 & 0\ftn[2]     &              &        \\
wCdS              &     0.93         & 1.98        & 2.24     &       &  2.50  & 0.03         & 0.07         & 2.60   \\
SrTiO$_3$         &     1.76         & 3.83        & 4.54     & 3.59  & $\sim$3.3&            &              &        \\
ScN               &    -0.26         & 0.95        & 1.24     & 0.96  & $\sim$0.9&0.01\ftn[2]            &              &        \\
NiO               &     0.45         & 1.1         & 1.6      &       &  4.3   &              &    &               \\
Cu\ftn[4]         & \mn2.33          & \mn2.35     & \mn2.23  &\mn2.18& \mn2.78&              &    &               \\
Cu\ftn[5]         & \mn2.33          & \mn2.85     & \mn2.73  &\mn2.18& \mn2.78&              &    &               \\
Gd$^{\uparrow}$   &  \mn4.6          & \mn5.6      & \mn6.2   &\mn4.1 & \mn7.9 &              &    &               \\
Gd$^{\downarrow}$ &     0.3          &  0.2        & 1.8      & 1.5   &  4.3   &              &    &
\end{tabular}
\footnotetext[1]{See Ref.~\onlinecite{gapdiscrepancy}}
\footnotetext[2]{LDA calculation}
\footnotetext[3]{Estimated from ZnSe}
\footnotetext[4]{Position of $\Gamma_{12}$ $d$ level, with $\ef$ set to charge-neutral point}
\footnotetext[5]{Position of $\Gamma_{12}$ $d$ level, with $\ef$ set to LDA value}
\end{table}


The calculations in Table~\ref{tab:gladgaps} support the argument that
using $Z$=1 is a better approximation than including $Z$: semiconductor
bandgaps are in significantly better agreement with experiment.  They
continue to be smaller than experimental values, which can be qualitatively
understood as follows.  Using $Z$=1 corresponds to updating $G$, but
leaving $W$ determined from the LDA eigenfunctions and eigenvalues.
Because the gap is underestimated in the construction of $\Pi$ and $W$,
$\Pi$ is overestimated so that $W$ is screened too strongly; thus $\Sigma$
is too small.  It is interesting, however, that QPEs evaluated with $Z$=1
can be rather good at times because of a fortuitous cancellation of errors.
We can refine self-consistency by updating $W$ in a manner similar to the
updating of $G$: that is using eigenvalues from Eq.~(\ref{eq:e1shot}) in
the calculation of $\Pi$.  However, $\epsilon$ computed in the RPA
$(\epsilon=1-v\Pi)$, omits excitonic effects.  Inclusion of electron-hole
correlations to $\Pi$ (via e.g., ladder diagrams) increases
${\rm{Im}}\,\epsilon(\omega)$ for $\omega$ in the vicinity of the gap for
semiconductors.  There is a concomitant increase in
${\rm{Re}}\,\epsilon(\omega)$ for $\omega\to 0$, as is evident by the
Kramers-Kronig relations; see e.g., Ref.~\onlinecite{arnaud01}.  Errors
resulting from the neglect of excitonic contributions to $\epsilon$
partially cancel errors resulting from LDA eigenvalues, as shown by Arnaud
and Alouani\cite{arnaud01}.  Thus, $W$ calculated from LDA eigenvalues is
not so bad in many cases because of this cancellation.  Often
$\epsilon_{\infty}$ calculated from the LDA eigenvalues is better than
$\epsilon_{\infty}$ calculated from LDA eigenvalues shifted by a scissors
operator to match the experimental band gap (See TABLE III in
Ref.~\onlinecite{arnaud01}).  This cancellation means that $GW$($Z$=1) can
be often be rather good, since $W$ itself is also better than what would be
obtained from (eigenvalue) self-consistency.  Table~\ref{tab:gladgaps}
shows that the fundamental gap for $GW(Z=1)$ is quite good for
mostly-covalent semiconductors such as Si or GaAs, but that the agreement
deteriorates as the ionicity increases.

\subsection{Off-diagonal contributions of $\Sigma$}

The usual $GW$A in Eq.~(\ref{eq:e1shot}) does not include the off-diagonal
contribution of $\Sigma-V\xc^{\rm{LDA}}$.  A simple way to take into
account the contribution off-diagonal parts is to replace the
energy-dependent matrix $\Sigma$ with some static hermitian matrix $\vxc$
as in the following, and to solve the eigenvalue problem, replacing
$V\xc^{\rm{LDA}}$ in the LDA hamiltonian with this potential.  We take
\begin{eqnarray}
\vxc = \frac{1}{2}\sum_{ij} |\psi_i\rangle
       \left\{ {{\rm Re}[\Sigma(\ei)]_{ij}+{\rm Re}[\Sigma(\ej)]_{ij}} \right\}
       \langle\psi_j|,
\label{eq:veff}
\end{eqnarray}
for $\Sigma$. Here {\rm Re} signifies the hermitian part,
the eigenvalues $\ei$ and the eigenfunctions $\psi_i$ are in LDA.
This $\vxc$ is used in our QP self-consistent $GW$ method~\cite{faleev04,qp06,chantis06a}.
This $\vxc$ retains the diagonal part contribution as in Eq.~(\ref{eq:e1shot})
(we now consider $Z$=1 case). From the perspective of the QP self-consistent $GW$ method,
including the off-diagonal $\Sigma$ corresponds to the first iteration,
and the LDA corresponds to the 0th iteration.  Table \ref{tab:gladgaps}
shows how the fundamental gap is affected by the off-diagonal parts of
$\vxc$ for selected semiconductors. Because the semiconductor
eigenfunctions and density are already rather good, the off-diagonal
contributions are small.  Contributions from the off-diagonal part of
$\vxc$ significantly increase when eigenfunctions have significant $d$
character (see SrTiO$_3$ and ScN in Table~\ref{tab:gladgaps}).  For
correlated systems the effects can be rather dramatic; see Ref.~\onlinecite{qp06}
how the QPEs are affected by the off-diagonal parts of $\Sigma$ in CeO$_2$.

In general, $GW$A errors are rather closely tied to the quality of LDA
starting point.  In the covalent $sp$ semiconductors C, Si and Ge, $GW$
gaps are rather good for $Z$=1.  In the series Zn(Te,Se,S,O), the deviation
between the LDA and experimental gap steadily worsens, and so does the $GW$
gap.  For ZnO, and even more so in CuBr, the $GW$ gap falls far below
experiment.  For these simple $sp$ materials, errors are related to their
ionicity, which can be seen qualitatively as follows.  As ionicity
increases, the dielectric response becomes smaller; consequently the
nonlocality missing from the LDA exchange-correlation
potential\cite{Maksimov89} becomes progressively more important.  Roughly
speaking, a reasonable picture of electronic structure in $sp$ systems
resembles an interpolation between the LDA, which has no nonlocality in the
exchange and underestimates gaps, and Hartree-Fock, which has nonlocality
but wildly overestimates gaps because the nonlocal exchange is not
screened.  As ionicity increases the gap widens and the dielectric function
decreases.  As the screening is reduced the LDA becomes a progressively
worse approximation.  Thus, the LDA is not an adequate starting point for
$GW$ in the latter cases.



Discrepancies between $GW$ and experiments become drastic when electronic
correlations are strong.  The $GW$A band gap for the antiferromagnetic-II
NiO is far from experiment, and moreover the conduction band minimum falls
at the wrong place (between $\Gamma$ and X \cite{faleev04}).
As Table~\ref{tab:gladgaps} shows, the LDA puts $f$ levels in Gd too close
to $\ef$. $GW$A results are only moderately better: shifts in the Gd $f$
level relative to the LDA are severely underestimated (see Table
~\ref{tab:gladgaps}).

$GW$ based on the LDA fails even qualitatively in CoO: it predicts a metal
with $\ef$ passing through an itinerant band of $d$ character.  In this
case, the $GW$A gives essentially meaningless results.  To get reasonable
results it is essential to apply the $GW$A with a starting point that
already has a gap. To get band gap for CoO in the single band picture, a
non-local potential which breaks time-reversal symmetry is required,
something which is not built into the local potential of the LDA.  Similar
problems occur with ErAs: the LDA predicts a very narrow minority $f$ band
straddling $\ef$, whereas in reality the minority $f$ manifold is
exchange-split into several distinct levels well removed from
$\ef$\cite{Komesu03}.  $GW$A shifts the minority $f$ levels only slightly
relative to the LDA: the entire $f$ manifold remains clustered in a narrow
band at the Fermi level, appearing once again qualitatively similar to the
LDA.

Generally speaking, the $GW$A (even $GW$A(Z=1)) is reasonable only under
limited circumstances---when the LDA itself is already reasonable.

%
%


\subsection{Band disentanglement problem}

\begin{figure}[htbp]
\centering
\includegraphics[angle=0,width=.45\textwidth,clip]{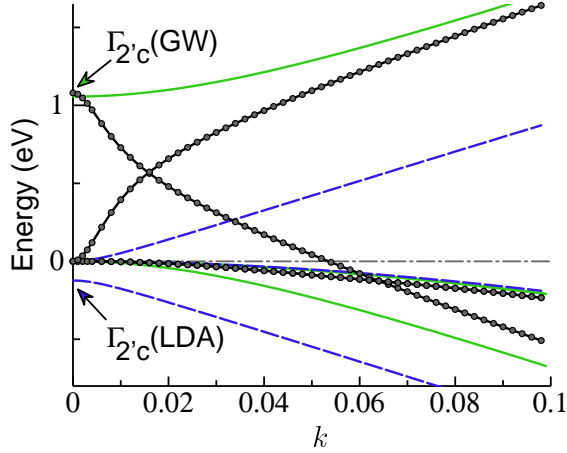}
\caption{Energy bands in Ge for ${\bf k}= 2\pi/a\,[00k]$ for small $k$
  within the $GW$A, using $Z$=1.  Spin orbit coupling was omitted.  Three
  approximations are compared: LDA (dashed blue line), $GW$ in the
  diagonal-$\Sigma$ only approximation, Eq.~(\ref{eq:e1shot}) (black line
  with circles), and $GW$ with $\Sigma$ computed according to
  Eq.~(\ref{eq:veff}) (solid green line).  In all three cases the three
  states of $p$ character ($\Gamma_{25'}$ symmetry) form the valence band
  maximum; this was taken to be the energy zero.  The LDA predicts the
  conduction band, the $\Gamma_{2'}$ state of $s$ character, to be slightly
  negative, causing the energy bands to be wrongly ordered at $\Gamma$.
  For $k>0$, the $\Gamma_{25'}$ state of $p_z$ symmetry couples to the
  $\Gamma_{2'}$ state, and the two repel each other.  Both kinds of $GW$
  put $\Gamma_{2'c}$ at approximately the correct position, 1~eV.  However,
  the diagonal-only $GW$ must follow the topology of the LDA: the
  eigenvectors are unchanged from the LDA.  Therefore the band starting out
  at $\Gamma_{2'c}$ sweeps downward, while the $p_z$ band starting out at
  $\Gamma_{25'v}$ sweeps upward, and the two bands cross near $k=0.02$.
  When the off-diagonal parts of $\Sigma$ are included, these two bands
  repel each other as they should.}
\label{fig:bands-ge-smallk}
\end{figure}

Even for the simple $sp$ semiconductors, there can be a ``band
disentanglement problem'' as a consequence of the diagonal-only
approximation.  At times the LDA orders energy levels wrongly: in hcp Co,
for example, it inverts the order of the minority $\Gamma_5$ and $\Gamma_3$
levels, which correspond to states of $L_3$ and $L_{2'}$ symmetry in the
fcc structure.  Wrong ordering of levels is a particularly serious
difficulty for narrow-gap semiconductors such as Ge, InAs, InSb, and InN.
Because the LDA underestimates bandgaps, the energy band structure around
$\Gamma$ has an inverted structure: the $s$-like conduction band of
$\Gamma_1$ symmetry (labeled as $\Gamma_{2'}$ in the homopolar case)
incorrectly falls below the $p$-like states of $\Gamma_{25'}$ symmetry.

When the $GW$A is evaluated from Eq.~(\ref{eq:e1shot}), the energy bands
retain the same connectivity as in the LDA, as
Fig.~\ref{fig:bands-ge-smallk} shows.  Consequently the conduction band has
a nonsensical negative mass near $\Gamma$, and it crosses with one of the
valence bands.  The diagonal-only approximation cannot make Ge an insulator
in principle, even though the levels are properly ordered at $\Gamma$.
This problem is avoided if the off-diagonal parts of $\Sigma$ are included,
as Fig.~\ref{fig:bands-ge-smallk} shows.  The conduction-band effective
mass in the latter case is computed to be $m^*=0.042m_0$, in good agreement
with a value of $m^*=0.038m_0$ estimated from magnetopiezoreflectance
spectra\cite{Aggarwal70}.  This shows that the off-diagonal contributions
of $\Sigma$ are reasonably well described by Eq.~(\ref{eq:veff}).

\section{Conclusions}
\label{sec:conclusions}

To conclude, we have analyzed various possible sources of error in
implementations of the \emph{GW}A, using calculations based on an
all-electron method with generalized Linear Muffin Tin Orbitals as a basis.
We analyzed convergence in QPEs with the number unoccupied states $N'$: the
rate of convergence for intermediate $N'$ (where the LMTO energy bands were
shown to precisely replicate APW bands), was qualitatively similar to, but
roughly twice that of a PP analysis by Tiago, Ismail-Beigi, and
Louie\cite{tiago04}.  On the other hand, it closely tracked the convergence
calculated by an LAPW+local orbitals method, which had a very similar LDA
band structure.  More generally, our $GW$A that properly
subtracts $V_{\rm{}xc}^{\rm{}LDA}$ calculated from the full density are in
reasonable agreement with each other
\cite{kotani02,weiku02,faleev04,Usuda04,Christoph06}; those that subtract
valence density only\cite{hamada90,alouani03} are also in reasonable
agreement for cases such as Si and SiC where the cores are sufficiently
deep.  Our own experience suggests that the LDA
treatment of core levels, where QPE are computed from
Eq.~(\ref{eq:qpgwnc}), will be problematic for $GW$\cite{kotani02} unless
the cores are very deep.  Since a PP construction is an approximation whose
justification is grounded in an all-electron theory, we should expect $GW$
calculations based on an LDA PP should be similarly problematic.  There is
apparently a significant dependence on how cores are treated in PP
implementations\cite{Marini02,Rohlfing95,Fleszar05,Shirley97}, even in Si
and Cu with their deep 2$p$ and 3$p$ cores.

We then presented a new analysis of convergence that is of particular
importance for minimal-basis implementations, and argued that measuring
convergence in the traditional cutoff procedure---by number of unoccupied
states $N'$ as given in Figs.~\ref{fig:gapconvergence} and
\ref{fig:gapvsinversen}---are not particularly meaningful for a minimal
basis.  We presented an alternative truncation of the full Hilbert space of
eigenfunctions, namely to use the entire hilbert space of a relatively
small basis.  We showed that a suitably constructed minimal basis is
sufficient to precisely describe the $GW$A QPE within 1 Ry or so of the
Fermi level, and that this kind of cutoff procedure seems to be more
efficient than the traditional $N'$ cutoff of a large basis.  We also
showed that traditional linearization of basis functions, either explicit
in an all-electron method or implicit through the construction of a
pseudopotential, result in errors approximately independent of the size of
basis.  Addition of local orbitals to extend the linear approximation
results in modest shifts in $sp$ nitride and oxide compounds, and shifts of
order 1-2~eV in transition-metal oxides.

We analyzed core contributions to the self-energy, and showed that an
exchange-only treatment of the core is adequate in most cases.  For all but
the most shallow cores (such as Na $2p$ and Ga $3d$), we showed that it is
sufficient to include the core contribution to the polarization only; an
approximate and rather painless implementation was suggested.  These
results can provide a framework for improved treatment of the core within a
pseudopotential approximation.

Finally, we considered the adequacy of $GW$A based on the LDA, for
different kinds of materials, and also Eq.~(\ref{eq:e1shot}) as an
approximation to the $GW$A.  We presented logical and numerical
justifications that using $Z$=1 generally gives better band gaps in
insulators.  In general inclusion of the off-diagonal part of $\Sigma$ and
some kind of self-consistency is essential to make the $GW$A a universally
applicable and predictive tool.  Taking into account both theoretical and
practical aspects, the quasi-particle self-consistent $GW$ scheme we have
proposed~\cite{faleev04,qp06,chantis06a} has the potential to be an
excellent candidate for such a tool: it obviates some of the difficulties
seen in the standard self-consistency, it no longer depends on the LDA, and
it appears to predict QPEs in a consistently reliable way for broad classes
of materials.

This work was supported by ONR contract N00014-02-1-1025.  S.~F. was
supported by DOE, BES Contract No. DE-AC04-94AL85000.  We thank the Ira
A. Fulton High Performance Computing Initiative for use of their computer
facilities.

\appendix
\section{Justification for $Z=1$.}
\label{appendix:z1}

Let us consider a limited self-consistency within $GW$A as follows.
We restrict self-consistency as follows:
\begin{enumerate}
\item We make only the QPE self-consistent.  Eigenfunctions are constrained
      to be the LDA eigenfunctions.
\item $W$ is assumed to be fixed.  Thus only the eigenvalues entering into
      $G$ are made self-consistent.
\end{enumerate}
%

Under these assumptions, we can show that QPE are rather well approximated
by Eq.~(\ref{eq:e1shot}) with $Z$=1.  To illustrate it, consider a
two-states model whose LDA eigenvalues and eigenfunctions are given by
$\psi_1,\varepsilon_1$, and $\psi_2,\varepsilon_2$, and the Fermi energy
falls between these states: $\varepsilon_2>E_{\rm F}>\varepsilon_1$.  Then
the LDA Green's function is
\begin{eqnarray}
G^{\rm{LDA}}(\omega) = \frac{|\psi_1\rangle\langle\psi_1|}{\omega -
            \varepsilon_1 -i\delta} +
            \frac{|\psi_2\rangle\langle\psi_2|}{\omega - \varepsilon_2
            +i\delta}.
\end{eqnarray}
After the limited self-consistency is attained, we will have
eigenvalues:
\begin{eqnarray}
G(\omega) =   \frac{|\psi_1\rangle\langle\psi_1|}{\omega - E_1 -i\delta}
            + \frac{|\psi_2\rangle\langle\psi_2|}{\omega - E_2 +i\delta},
\label{eq:gsc}
\end{eqnarray}
where $E_1$ is given by
\begin{eqnarray}
E_1 = \varepsilon_1 + {\rm Re}\langle\psi_1|\Sigma(E_1,[G])- V^{\rm{LDA}}_{\rm{xc}}|\psi_1\rangle .
\end{eqnarray}
There is a similar equation for $E_2$.
Note that $\Sigma(E_1,[G])$ is calculated in $GW$A from $G$ of
Eq.~(\ref{eq:gsc}) at $E_1$.

As we can expect that $W$ is dominated by diagonal terms
$W_1(\omega)=\langle\psi_1 \psi_1|W(\omega) |\psi_1 \psi_1\rangle$ and
$W_2(\omega)=\langle\psi_2 \psi_2|W(\omega) |\psi_2 \psi_2\rangle$, we
neglect other matrix elements of $W(\omega)$.  Then $\Sigma$ becomes
\begin{eqnarray}
{\rm Re} \langle\psi_1|(\Sigma(E_1,[G])|\psi_1\rangle
&=& {\rm Re} \int \langle\psi_1| i G(E_1+\omega') W(\omega')|\psi_1\rangle \,d \omega'\nonumber\\
&\approx& {\rm Re} \int \frac{iW_1(\omega') \,d \omega'}{E_1 + \omega' - E_1 -i\delta} \nonumber\\
&=& {\rm Re} \int \frac{iW_1(\omega') \,d \omega'}{\varepsilon_1 + \omega' - \varepsilon_1 -i\delta} \nonumber\\
&=& {\rm Re} \langle\psi_1|\,\Sigma\left(\varepsilon_1,[G^{\rm{LDA}}]\right)\,|\psi_1\rangle
\label{eq:sigmawithoutz}
\end{eqnarray}
A similar equation applies for $E_2$.  The energy shift $E_1\to{}\varepsilon_1$
entering into the evaluation $\Sigma$ is exactly compensated by the energy
shift in $G\to{}G^{\rm{LDA}}$.  Or equivalently using $Z$=1 is an
approximate way to obtain self-consistency.  Eq.~(\ref{eq:sigmawithoutz})
corresponds Eq.~(\ref{eq:e1shot}) with $Z$=1.



\begin{thebibliography}{51}
\bibitem[{gap({\natexlab{a}})}]{gapnote}
\bibinfo{note}{The $GW$ fundamental gap for Si was incorrectly quoted as
  0.84~eV in Ref.~\onlinecite{faleev04}. The correct value is 0.91~eV.}
\bibitem[{ltr()}]{ltrunc}
\bibinfo{note}{In the LDA a small amount of the envelope function remains
  inside the augmentation spheres, to handle incomplete $l$ convergence of the
  augmentation functions. In the present $GW$ implementation this feature is
  not retained, because of the nonlocal character of the potential. As a
  consequence wave function products must be augmented to higher $l$ than in
  the LDA case.}
\bibitem[{cor({\natexlab{a}})}]{coretrunc}
\bibinfo{note}{In the $GW$ scheme as we have currently implemented it, the LDA
  core states are approximated by truncating them outside the MT spheres, and
  scaling the part inside the MT sphere to conserve charge. An alternatively
  would be to integrate the core wave function subject to the boundary
  condition that the value and slope vanish at the MT boundary. In either case,
  the core is not strictly orthogonal to the valence; this appears to be the
  primary source of error in the present treatment of the core. Experience with
  the more precise local orbitals treatment shows that this approximation is
  not serious at the ``exchange-only'' level. However their contribution to
  correlation is reliable only for relatively deep cores such as the Si $2p$.}
\bibitem[{kco()}]{kconvergence}
\bibinfo{note}{Using the linear tetrahedron method, a mesh of
  $6\times{}6\times{}6$ divisions (doubling the number of points in the energy
  denominator) resulted in an approximately rigid shift of +0.02~eV relative to
  a calculation with an $8\times{}8\times{}8$ mesh, while a
  $4\times{}4\times{}4$ mesh was found to overestimate the conduction bands by
  0.10~eV. These shifts were approximately unchanged when varying, e.g., the
  basis set or the number of unoccupied states.}
\bibitem[{cor({\natexlab{b}})}]{corelocal}
\bibinfo{note}{We can treat the core state explicitly as valence states through
  the use of local orbitals, or solve for the core independently; see
  Sec.~\ref{subsec:LDA}. For a deep core such as the Si 2$p$ it matters little
  whether the core eigenfunctions are computed separately or generated through
  a local orbital, even with the approximate core treatment \cite{coretrunc}.
  This is expected since the core level is very deep. Since the LDA
  eigenfunctions are represented completely differently in the two cases, the
  stability is a confirmation of the robustness of the method. For shallow
  cores with their larger radial extent, approximations in our present
  treatment of core states\cite{coretrunc} do not enable us to distinguish true
  core contributions from artifacts of the implementation, so we use the local
  orbital approach here.}
\bibitem[{lmt()}]{lmtonote}
\bibinfo{note}{We reiterate the present LMTO method has a key advantage in that
  it can be made small, and also highly precise when needed; see
  Fig.~\ref{fig:sildabands}.}
\bibitem[{ntr()}]{ntrunc}
\bibinfo{note}{Those familiar with the usual $N'$ truncation analysis might ask
  why if applying it to a small-basis set calculation results in a different
  convergence rate, whether the small-basis calculation is adequately
  converged. This can be understood as follows. High-lying eigenvalues in the
  small basis tends to some kind of average of the true higher states; unlike a
  basis that spans the hilbert space of those states, individual states do not
  have physical meaning. Only the collective contribution from the sum over
  higher lying states to Eqs.~\ref{eq:polf} and \ref{eq:sigc} has physical
  meaning in a finite basis, namely $\Pi_{IJ}(\omega)$ and and
  $\Sigma_c(\omega)$ within the hilbert space of eigenfunctions. All that is
  necessary for completeness is that the basis have enough freedom in it to
  describe $\Pi_{IJ}(\omega)$ and $\Sigma_c(\omega)$ on the energy scales of
  relevance. The completeness of a particular basis must be checked numerically
  by making it ever-more complete and monitoring the convergence, as is done in
  traditional LDA calculations.}
\bibitem[{gap({\natexlab{b}})}]{gapdiscrepancy}
\bibinfo{note}{QP levels calculated by ourselves in Ref.~\onlinecite{kotani02}
  did not have local orbitals included. Mostly for that reason, the QP levels
  are a little different from those cited here. Also, the gap in wurtzite InN
  cited here is 0.17~eV larger than the 0.02~eV reported by
  Usuda\cite{Usuda04}, who used LAPW eigenfunctions that did not have local
  orbitals. (Leaving out the N $3p$ and In $5d$ local orbitals, we obtained a
  gap of 0.01~eV with an LMTO basis in Ref.~\onlinecite{kotani02}.) The basis
  sets used in the present calculations are substantially larger, and
  tolerances set tighter than in what Ref.~\onlinecite{kotani02} used. These
  changes make additional corrections of $\sim{}0.05$~eV. Finally the
  fundamental gap for Si was incorrectly tabulated as 0.84~eV in
  Ref.~\onlinecite{faleev04}: the actual value for the basis used in that paper
  was 0.91~eV.}
\end{thebibliography}

\end{document}